\RequirePackage{ifthen}

\newboolean{papertwocol}

\setboolean{papertwocol}{false}

\ifthenelse{\boolean{papertwocol}}{
    
}
{
    
}

\documentclass{article}

\usepackage{arxiv}

\usepackage[utf8]{inputenc} 
\usepackage[T1]{fontenc}    
\usepackage{hyperref}       
\usepackage{url}            
\usepackage{booktabs}       
\usepackage{amsfonts}       
\usepackage{nicefrac}       
\usepackage{microtype}      
\usepackage{lipsum}		
\usepackage{graphicx}
\usepackage{doi}

\usepackage{xparse}
\usepackage{amsmath}
\usepackage{multirow}
\usepackage{tikz}
\usepackage{url}
\usepackage{amsthm}
\usepackage{xspace}
\usepackage{caption}
\usepackage{subcaption}
\usepackage{algorithm}
\usepackage[noend]{algpseudocode}

\providecommand{\qfrdl}{QF\_RDL\xspace}
\providecommand{\qflra}{QF\_LRA\xspace}
\providecommand{\Tau}{\mathrm{T}\xspace}
\providecommand{\Tau}{\mathrm{T}\xspace}
\NewDocumentCommand\Rmax{oo}{%
        \IfNoValueTF{#1}
        {\mathbb{R}_{\max}}
        {
    \IfNoValueTF{#2}
    {\mathbb{R}_{\max}^{#1}}
    {\mathbb{R}_{\max}^{#1 \times #2}}
    }
}
\NewDocumentCommand\IRmax{oo}{%
        \IfNoValueTF{#1}
        {\mathbb{IR}_{\max}}
        {
    \IfNoValueTF{#2}
    {\mathbb{IR}_{\max}^{#1}}
    {\mathbb{IR}_{\max}^{#1 \times #2}}
    }
}
\newcommand{\ignore}[1]{{}}
\newcommand{\tuple}[1]{\langle #1 \rangle}
\newcommand*{\QEDB}{\hfill\ensuremath{\square}}
\newcommand{\CommentX}[1]{\mbox{}\hfill~#1~}

\newtheorem{defn}{Definition}

\newtheorem{exmp}{Example}
\newtheorem{prop}{Proposition}

\newcommand{\ltlX}{\ensuremath{\textbf{X}}\xspace}
\newcommand{\ltlF}{\ensuremath{\textbf{F}}\xspace}
\newcommand{\ltlG}{\ensuremath{\textbf{G}}\xspace}
\newcommand{\ltlU}{\ensuremath{\xspace\textbf{U}}\xspace}
\newcommand{\ltlR}{\ensuremath{\xspace\textbf{R}}\xspace}
\newlength{\hack}

\usetikzlibrary{arrows,patterns,positioning,intersections,calc,shapes}
\tikzstyle{decision} = [diamond, draw, text width=4.5em, text badly centered, node distance=3cm, inner sep=0pt,scale=0.75]
\tikzstyle{block} = [rectangle, draw, text width=5em, text centered, rounded corners, minimum height=2em,scale=0.75]
\tikzstyle{par} = [trapezium, draw, trapezium left angle=75, trapezium right angle=-75, inner sep=1pt, trapezium stretches=true, trapezium stretches body=true, text width=8em, text centered,minimum height=2.5em,scale=0.75]
\tikzstyle{line} = [draw, -latex']
\title{Formal Analysis and Verification
of Max-Plus Linear Systems}

\author{\href{https://orcid.org/0000-0003-0817-1106}{\includegraphics[scale=0.06]{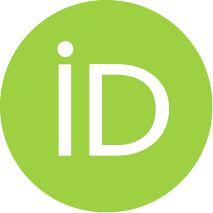}\hspace{1mm}Muhammad Syifa'ul Mufid} \\
	Department of Mathematics\\
	Institut Teknologi Sepuluh Nopember\\
	Surabaya, Indonesia \\
	\texttt{syifaul.mufid@matematika.its.ac.id} \\
	\And
	\href{https://orcid.org/0000-0002-6370-1061}{\includegraphics[scale=0.06]{orcid.pdf}\hspace{1mm}Andrea Micheli} \\
	Fondazione Bruno Kessler\\
    Trento, Italy\\
	\texttt{amicheli@fbk.eu} \\
    \And
    \href{https://orcid.org/0000-0002-5627-9093}{\includegraphics[scale=0.06]{orcid.pdf}\hspace{1mm}Alessandro Abate} \\
	Department of Computer Science\\
    University of Oxford\\
    Oxford, United Kingdom\\
	\texttt{alessandro.abate@cs.ox.ac.uk} \
    \And
    \href{https://orcid.org/0000-0002-1315-6990}{\includegraphics[scale=0.06]{orcid.pdf}\hspace{1mm}Alessandro Cimatti} \\
	Fondazione Bruno Kessler\\
    Trento, Italy\\
	\texttt{cimatti@fbk.eu} \
}

\begin{document}
\maketitle
\begin{abstract}
Max-Plus Linear (MPL) systems are an algebraic formalism with practical applications in transportation networks, manufacturing and biological systems.
In this paper, we investigate the problem of automatically analyzing the properties of MPL,
taking into account both structural properties such as transient and cyclicity, and the open problem of user-defined temporal properties.
We propose Time-Difference LTL (TDLTL), a logic that encompasses the delays between the discrete time events governed by an MPL system, and characterize the problem of model checking TDLTL over MPL. 
We first consider a framework based on the verification of infinite-state transition systems, and propose an approach based on an encoding into model checking.
Then, we leverage the specific features of MPL systems to devise a highly
optimized, combinational approach based on Satisfiability Modulo Theory (SMT).
We
experimentally evaluate the features of the proposed approaches on a large set of benchmarks.
The results show that the proposed approach substantially outperforms the state of the art competitors in expressiveness and effectiveness, and demonstrate the superiority of the combinational approach over the reduction to model checking.
\end{abstract}
\section{Introduction}


Max-Plus Linear (MPL) systems are a class of discrete-event dynamic systems (DEDS) based on the Max-Plus algebra, an algebraic structure that uses maximisation and addition as its binary operations. MPL systems are employed to model processes with features of synchronization but without concurrency, and as such are widely used for applications in transportation networks \cite{Baccelli}, manufacturing \cite{Heidergott} and biological systems \cite{Chris,Comet}. In MPL models, the states correspond to time instances related to discrete events.


In this paper, we tackle the general problem of automatically analyzing the properties of MPL behaviours over time.
%
%
We first consider the fundamental and well-studied \emph{structural property} of MPL systems related to periodic behaviors. Given an initial vector, the trajectories of an MPL system are eventually periodic (in max-plus algebraic sense), starting from a specific bound called the \textit{transient}, and with a specific period called \textit{cyclicity} \cite{Baccelli}.
As explained in \cite[Section 3.1]{Heidergott}, the transient is closely related to the notion of \textit{cycle-time} vector, which governs the asymptotic behaviour of MPL systems.
%
Such a bound is key to solve a number of fundamental problems of MPL systems, namely reachability analysis \cite{Mufid2020} and bounded model checking \cite{Mufid2019}:
it plays a crucial role as the ``completeness threshold'' (namely, the maximum iteration that is sufficient for the termination of the algorithm) \cite{CT} for those two problems. On the one hand, the cyclicity is directly related to the precedence graph of MPL systems and its computation is relatively easy.
On the other hand, the transient is hard to compute and in general not correlated to the dimension of the MPL system. Thus, it is possible for the resulting transient to be relatively large for a small-dimensional MPL system.
There are several known upper bounds \cite{Charron,Merlet,Nowak,Gerardo} for the transient, which are mostly computed via the corresponding precedence graph and are, in practice, much larger than the actual values. Hence, the ability to compute transients is important in practice.

%
In this work we newly consider the problem of formally specifying and analyzing \emph{user-defined} temporal properties or specifications for MPL systems. Properties of interest include, for example, whether the difference between two time-stamps of discrete-events is always bounded, or if the increment between different time steps is constant.
Although this problem is important for the analysis of MPL systems, somewhat surprisingly it is largely ignored by the existing literature.\ignore{WAS:
This problem is of paramount importance for the analysis of MPL systems and is largely ignored by existing literature.}



\paragraph*{Contributions}

In this paper, we make the following contributions. First, we define \textit{time-difference} LTL (TDLTL), a logic to express complex temporal properties of MPL systems, and formalize the problem of model checking TDLTL specifications over MPL.
TDLTL combines temporal operators that are typical of Linear Temporal Logic (LTL), with basic atoms that relate the values of the MPL system variables over different time points, hence encompassing the delays between  discrete-time events.

Second, we show that MPL systems can be encoded as infinite-state transition systems, which can then be analyzed by existing model checking procedures. In particular, we target symbolically-represented transition systems, for the analysis of both structural and user-defined properties of MPL. Our encoding is such that every execution of the input MPL system is simulated by a trace of the symbolic transition system and vice versa, allowing to tackle both the analysis of structural properties and the model-checking of TDLTL formulae.


Third, we describe specialized procedures based on the use of a symbolic representation in Satisfiability Modulo Theory (SMT) solving~\cite{Barret}, leveraging the properties of MPL systems. We provide a novel, SMT-based procedure to compute the transient and cyclicity of MPL systems. The main idea is to transform the problem instance into a formula in Quantifier-Free Real Difference Logic (\qfrdl) \cite{Cotton}, that is a Boolean combination of atoms in the form $v_i - v_j \bowtie c$ where $\bowtie \hspace{0.5ex}\in \{>, <, \ge, \le, =\}$, $v_i$ is a real variable, and $c$ a rational constant. An SMT solver is then invoked on a series of \qfrdl formulae to compute the transient and cyclicity.
%
%
We also propose a family of SMT-based, specialized algorithms for TDLTL model
checking of MPL systems.
The algorithms, which we prove to be sound and complete, leverage the
periodic behaviour of an MPL system: intuitively, the transient and
cyclicity of the MPL system induce a completeness threshold for a
bounded encoding of the verification problem.
The family of algorithms has several variants, depending on two
independent factors.  One is the \emph{computation of the transient bound}, that
could be carried out either upfront (i.e., before calling the SMT solver), or
incrementally, interleaving it with multiple solver calls.  The other
is the \emph{unrolling of the transition relation}, that can either
follow the explicit approach of Bounded Model Checking (BMC) \cite{Biere3}, or -- thanks to some algebraic
properties of MPL systems -- be left implicit, so that the number of
variables that the SMT solver must handle is significantly reduced. It should be noted that the resulting algorithms rely on the existence of transients: indeed, the verification problem for MPL systems that do not yield finite transients remains an open problem.

\paragraph*{Novelty}

Traditional analysis of the dynamics of MPL systems focuses on their
algebraic and graph representation (cf.~\autoref{precedence}), which
allows for the investigation of several structural problems, such as
eigenproblems~\cite{Cuninghame}, optimisation~\cite{Bouquard} and
periodic behaviour~\cite{Bart2,Nowak}.
Unlike the mentioned standard results,
the newer line of work in in~\cite{DiekyBackward,DiekyForward,Dieky2} deals with
reachability analysis and formal verification of  MPL systems. These methods are in general not complete, as they rely on the construction of an abstraction that overapproximates the concrete MPL system~\cite{Dieky1,Mufid2018,Mufid2019}. Furthermore, the abstraction procedures suffer from state-explosion problems, given that the size of the abstraction is exponential in the number of variables of the MPL, and are practically unable to deal with more than few variables.

Compared to these works, this paper has two key points of novelty: first, it provides the only general specification language to express properties of MPL systems;  second, it adopts symbolic techniques from the field of formal verification to achieve efficiency.

\paragraph*{Experimental evaluation}

The approach has been implemented and the algorithms have been experimentally evaluated on a large set of benchmarks. The results show that the proposed techniques can be applied to general TDLTL formulae on large MPL systems, the analysis of which is completely out-of-reach of existing  abstraction-based techniques~\cite{Mufid2019,Mufid2018}. The
comparison also shows that the SMT-based algorithms yield
orders-of-magnitude speed-up against the
{translational approach into symbolic transition system described in Section 5.}
%

\paragraph*{Structure of the paper}

In Section~\ref{sec-preliminaries} we provide the logical background on SMT and SMT-based model checking, while
Section~\ref{sec-MPLS} describes the basics of MPL systems. In Section~\ref{sec-problem-definition} we present the verification problems we tackle together with the novel TDLTL logic we use to express temporal properties.
In Sections~\ref{sec-algorithms-model-checking} and~\ref{sec-algorithms-smt} we present the verification algorithms based on reduction to model checking and on SMT, respectively.
In Section~\ref{sec-relatedwork} we discuss the relationship with related works.
Section~\ref{sec-experimental-evaluation} experimentally evaluates the algorithms and Section~\ref{sec-conclusions} draws the conclusions and presents directions for future work. \ifthenelse{\boolean{papertwocol}}{\textcolor{red}{\bf TODO: cite arXiv version here}}{{In Appendix~\ref{sec-case-study} we present the max-plus model of the London Underground network and also showcase the formal verification procedure using SMT-based algorithms. The additional experiments to compute transient and cyclicity are presented in Appendix~\ref{sec-more experiments}}}


\section{Preliminaries}
\label{sec-preliminaries}

In this section, we introduce the main concepts and notation for SMT and transition systems as well as some notions of symbolic model-checking. 

\subsection{Satisfiability Modulo Theory}

We consider the framework of first-order logic. A term is either an individual variable, a constant, or the application of an $n$-ary function symbol to $n$ terms; an atom is either a Boolean variable, or the application of an $n$-ary relation symbol to $n$ terms. A formula is either an atom, the application of negation ($\neg$) to a formula, the application of a binary Boolean connective to two formulae, i.e. conjunction ($\wedge$), disjunction ($\vee$), implication ($\rightarrow$), double implication ($\leftrightarrow$), or the application of a universal ($\forall$) or existential ($\exists$) quantifier to a variable and a formula. 
We assume the standard notions of free/bound variable, substitution, assignment/interpretation, model, truth, validity~\cite{HuthRyan}.

In Satisfiability Modulo Theory (SMT), the interpretation may be restricted to a given background theory of interest $\Tau$.
A widely used theory is Quantifier-Free Linear Real Arithmetic (\qflra). A formula in \qflra may contain the following theory symbols: constants in $\mathbb{Q}$, function symbols in $+,\cdot$, relational symbols in $\{>, <, \ge, \le, \ne, =\}$.
The \qflra theory atoms have the form $\sum_i a_i \cdot v_i \bowtie c$ where $\bowtie$ is a relational operator, every $v_i$ is a real variable, and every $a_i$ and $c$ are rational constants.
In \qflra, theory symbols are interpreted according to standard arithmetic.
Also relevant for this paper is Quantifier-Free Difference Logic (\qfrdl), the subset of \qflra in which all atoms are restricted to the form $v_i - v_j \bowtie c$. Both \qflra and \qfrdl are decidable, as they can geometrically express Boolean operations on octagons and convex polyhedra, for which decision procedures exist \cite{SMT}. 
%
Other practically useful theories, not interest for this work, include Nonlinear Real Arithmetic (NRA), Linear and Nonlinear Integer Arithmetic (LIA, NIA), Equality and Uninterpreted Functions (EUF), and the theory of arrays, and their combinations \cite{SMT}.

Given a first-order formula $\psi$, the SMT problem consists in deciding whether there exists a $\Tau$-model (i.e.\ an assignment to the free variables in $\psi$) that satisfies $\psi$ according to theory $\Tau$ \cite{SMT}.
For example, consider the formula $ (x \le y) \wedge (x + 3 = z) \vee (z \ge y) $ within the theory of real numbers.
The formula is satisfiable and a valid model is $\{x:=5, \: y:=6, \: z:=8\}$. SMT has grown into a very active research subject: it has a standardised library and also a collection of benchmarks developed by the SMT community \cite{SMT-LIB}, as well as an annual international competition for SMT solvers \cite{SMT-COMP}.  As a result, there are several powerful SMT solvers,  such as MathSAT5 \cite{MATHSAT}, Yices 2.6 \cite{Yices}, and Z3 \cite{Z3}, that can support different a variety of theories and their combinations~\cite{Daniel}. 
Applications of SMT solving arise on supervisory control of discrete-event systems \cite{Shoaei}, verification of neural networks \cite{Reluplex}, optimization \cite{Yi}, and beyond.

\subsection{Symbolic Transition Systems}
We consider symbolic infinite-state transition systems (STS). An STS $M$ is defined as a tuple of the form $\langle V, I(V), T(V,V')\rangle$.
$V$ is the set of state variables, where a variable may be either Boolean or real-valued.
A state is an assignment to $V$, i.e., a mapping from each variable $v_i$ to the corresponding range.
As standard in symbolic verification, a formula $\phi(V)$ is used to represent the set of (states corresponding to) its models, i.e. $\{\mu | \mu\models\phi\}$.
$I(V)$ is the set of initial states of $M$.
In order to represent transitions, we use a new set $V'=\{v'|v\in V\}$ of next-state variables. A transition is an assignment to $V\cup V'$, where the values to the $V$ and $V'$ variables represent the state before and after the transition, respectively.
The set of transitions of $M$ is the  set of assignments to $V\cup V'$ satisfying $T(V,V')$.

In the following, we assume that an STS $M=\langle V, I(V), T(V,V')\rangle$ is given, with $I$ and $T$ being formulae in \qflra. Intuitively, an STS models the behavior of the variables in $V$ by constraining the set of possible initial values to be one of the models of $I(V)$ and by enforcing the satisfaction of the formula $T(V,V')$ at each transition step. More formally, we denote states with $s, s_0, s_1, \ldots$, and we write $s(v)$ for the values assigned to $v$ in $s$.
We write $s'$ to denote the assignment to the $V'$ variables such that $s'(v')=s(v)$. 
A path is a sequence of states $s_0, s_1, \ldots, s_n$ such that $s_0$ is an initial state, i.e. $s_0\models I$, and for each $i$ there is a transition between $s_i$ and $s_{i+1}$, i.e. $s_i\cup s'_{i+1}\models T$.
A state $s$ is reachable iff there exists a path containing it.

STSs are very powerful and common representation tools in formal verification, in particular for the problem of model-checking, which consists in checking if a given STS satisfies a given property expressed in a (usually temporal) logic. In this paper, we concentrate on Linear Temporal Logic (LTL) model-checking for which we give the relevant background in the following.

\subsection{Linear Temporal Logic}

Let $V$ be a set of Boolean and real-valued variables.
We consider Linear Temporal Logic (LTL) over $V$, defined as follows:
\ifthenelse{\boolean{papertwocol}}{
\begin{align}
\begin{array}{cc}
\varphi:= & \top~|~\bot~|~p~|~\neg\varphi~|~\varphi_1 \wedge \varphi_2~|~\varphi_1 \vee \varphi_2~|~ \\
& \ltlX \varphi~|~\ltlF \varphi~|~\ltlG \varphi~|~\varphi_1 ~\ltlU~\varphi_2~|~\varphi_1 ~\ltlR~\varphi_2
\end{array}
\label{LTL_syntact}
\end{align}
}{
\begin{equation}
\varphi:=\top~|~\bot~|~p~|~\neg\varphi~|~\varphi_1 \wedge \varphi_2~|~\varphi_1 \vee \varphi_2~|~\ltlX \varphi~|~\ltlF \varphi~|~\ltlG \varphi~|~\varphi_1 ~\ltlU~\varphi_2~|~\varphi_1 ~\ltlR~\varphi_2
\label{LTL_syntact}
\end{equation}
}
where $p$ is a either a Boolean variable or a \qflra atom over the real-valued variables in $V$.
The temporal operators are \emph{next state} (\ltlX),
\emph{eventually} ($\ltlF$), \emph{globally} ($\ltlG$), \emph{until} ($\ltlU$) and \emph{release} ($\ltlR$).

An LTL formula $\phi(V)$ is interpreted over an infinite sequence $\pi$ of assignments to its variables $X$. We indicate with $\pi[i]$ the $i$-th assignment in $\pi$. Furthermore, $\pi[i..]$ is the $i$-th suffix of $\pi$.
The semantics of LTL is as follows.
\ifthenelse{\boolean{papertwocol}}{%
\begin{equation}
\left.
\label{LTL_semantic}
\begin{array}{ll}
\pi\models \top,&\\
\pi\not\models \bot,&\\
\pi\models p& \text{iff}~ \mbox{$p$ is satisfied in \qflra by the}\\
            & \quad\: \mbox{assignment $\pi[0]$},\\
\pi\models  \neg \varphi& \text{iff}~ \pi\not\models \varphi,\\
\pi\models \varphi_1\wedge \varphi_2& \text{iff}~\pi\models \varphi_1\text{ and } \pi\models\varphi_2,\\
\pi\models \varphi_1\vee \varphi_2& \text{iff}~\pi\models \varphi_1\text{ or } \pi\models\varphi_2,\\
\pi\models \ltlX \varphi& \text{iff}~\pi[1..]\models \varphi,\\
\pi\models \varphi_1 \ltlU\varphi_2& \text{iff}~\exists j\geq 0.~\pi[j..]\models \varphi_2~\text{and}\\
            & \quad\: \forall 0\leq i< j.~\pi[i..]\models \varphi_1,\\
\pi\models \varphi_1 \ltlR\varphi_2& \text{iff}~\forall j\geq 0.~\pi[j]\models \varphi_2~\text{or}~\exists i\geq 0.\\
            & (\pi[i..]\models \varphi_1\wedge \forall h\leq i. \pi[h..]\models \varphi_2 ),\\
 \pi\models  \ltlF \varphi& \text{iff}~\exists j\geq 0.~\pi[j..]\models \varphi,\\
 \pi\models \ltlG \varphi&\text{iff}~\forall j\geq 0.~\pi[j..]\models \varphi.
\end{array}\!\!\!
\right\}
\end{equation}
}{%
\begin{equation}
\left.
\label{LTL_semantic}
\begin{array}{ll}
\pi\models \top,&\\
\pi\not\models \bot,&\\
\pi\models p& \text{iff}~ \mbox{$p$ is satisfied in \qflra by the assignment $\pi[0]$},\\
\pi\models  \neg \varphi& \text{iff}~ \pi\not\models \varphi,\\
\pi\models \varphi_1\wedge \varphi_2& \text{iff}~\pi\models \varphi_1\text{ and } \pi\models\varphi_2,\\
\pi\models \varphi_1\vee \varphi_2& \text{iff}~\pi\models \varphi_1\text{ or } \pi\models\varphi_2,\\
\pi\models \ltlX \varphi& \text{iff}~\pi[1..]\models \varphi,\\
\pi\models \varphi_1 ~\ltlU~\varphi_2& \text{iff}~\exists j\geq 0.~\pi[j..]\models \varphi_2~\text{and}~\forall 0\leq i< j.~\pi[i..]\models \varphi_1,\\
\pi\models \varphi_1 ~\ltlR~\varphi_2& \text{iff}~\forall j\geq 0.~\pi[j..]\models \varphi_2~\text{or}~\exists i\geq 0.~(\pi[i..]\models \varphi_1\wedge \forall h\leq i.~ \pi[h..]\models \varphi_2 )\\
 \pi\models  \ltlF \varphi& \text{iff}~\exists j\geq 0.~\pi[j..]\models \varphi,\\
 \pi\models \ltlG \varphi&\text{iff}~\forall j\geq 0.~\pi[j..]\models \varphi.
\end{array}\!\!\!\right\}
\end{equation}
}
%

The following properties hold: $\phi~\ltlR~\psi \equiv \neg(\neg\phi~\ltlU~\neg\psi)$; $\ltlF\phi \equiv \top~\ltlU~\psi$; $\ltlG\phi\equiv \neg\ltlF\neg\phi$.

Given a STS $M$, we say that a LTL formula $\phi$ holds in $M$, written $M\models\phi$, iff for each infinite path $\pi$ of $M$, $\pi \models \phi$. Checking whether a given LTL formula holds in a given STS is the LTL model-checking problem.

\subsection{Symbolic Model Checking}


Depending on the assumptions on the STS $M$ and the temporal formula $\phi$, the model-checking problem varies in complexity and different techniques have been presented to handle different formulations of it. If $M$ only contains Boolean variables, the set of reachable states of $M$ is finite, and the model-checking problem is decidable. Over the years, many different approaches to symbolic model checking have been proposed: algorithms based on fix point computation~\cite{DBLP:books/daglib/0071856,DBLP:journals/iandc/BurchCMDH92} based on Binary Decision Diagrams~\cite{bryant}, and on Boolean satisfiability (SAT)~\cite{SATHandbook}, such as Bounded Model Checking (BMC)~\cite{Biere3}, Induction~\cite{Sheeran}, Interpolation~\cite{mcmillan}, and IC3~\cite{DBLP:conf/vmcai/Bradley11}.
When $M$ has an infinite state space, the problem is, in general, undecidable.
Despite this, several approaches exist, based on abstraction and SMT solving (e.g. IC3ia~\cite{ic3ia}).
%
%
Particularly interesting for this paper is BMC. BMC is a technique focused on finding violations to the given property, i.e. producing a trace $\pi$ that satisfies $\neg\phi$, hence witnessing a violation.
The idea is to encode the existence of such $\pi$ as a formula
\ifthenelse{\boolean{papertwocol}}{
\begin{align*}
I(V^{(0)}&)\wedge T(V^{(0)},V^{(1)})\wedge\ldots\wedge\\
& T(V^{(k-1)}, V^{(k)})\wedge T(V^{(k)}, V^{(l)}) \wedge[[\neg\phi]]^l_k
\end{align*}
}{
%
$$
I(V^{(0)})\wedge T(V^{(0)},V^{(1)})\wedge\ldots\wedge\\
T(V^{(k-1)}, V^{(k)})\wedge T(V^{(k)}, V^{(l)}) \wedge[[\neg\phi]]^l_k
$$
}
that is satisfiable if and only if there exists a finite path of $k+1$ transitions (represents an infinite trace) that violates $\phi$.
The set of state variables $V$ is replicated $k+1$ times so that $V^{(i)}$ represents the $i$-th state of the trace.
The constraint $T(V^{(k)}, V^{(l)})$ encodes the existence of a loopback, by stating that the successor of the last state $V^{(k)}$ is the previously visited $V^{(l)}$.
This makes it possible to finitely present a lasso-shaped path, i.e. $\pi=\alpha\cdot\beta^\omega$ for some finite sequences of states $\alpha$ and $\beta$.  
Finally, the formula 
$[[\neg \phi]]^l_k$ is a set of constraints over $V^{(0)}, \ldots,V^{(k)}$ that ensures that $\phi$ does not hold on $\pi$. 

In the finite-state case, BMC is guaranteed to find a violation if it exists, i.e. there exist $k,l$ such that $[[M\models\phi]]^l_k$ is satisfiable.
In the infinite-state instance, this is in general not the case: it is possible that the counterexamples do not have a ``looping structure'', i.e. can not be expressed as $\alpha\cdot\beta^\omega$.

Symbolic model checking is a computational problem of practical relevance and several tools exists. In this paper we use the \textsc{nuXmv} \cite{nuxmv} model-checker for our experimentation. \textsc{nuXmv} supports finite- and infinite-state STSs and implements many state-of-the-art algorithms for model checking of LTL formulae. 

\section{Max-Plus Linear Systems}
\label{sec-MPLS}


In this section, we present an overview of the main definitions and results concerning Max-Plus Linear Systems (abbreviated as MPL Systems throughout the rest of the paper). 

Max-plus algebra \cite{Baccelli} is a modification of the classical linear algebra, and is defined over the so-called max-plus semi-ring  $(\Rmax,\oplus,\otimes)$, where
$\Rmax:=\mathbb{R}\cup\{\varepsilon:=-\infty\}$ and 
\[
    a\oplus b := \max\{a,b\}, \qquad
    a\otimes b := a+b,
\]
for all $a,b\in \Rmax$. The zero and unit elements of $(\Rmax,\oplus,\otimes)$ are $\varepsilon$ and 0, respectively. By $\Rmax[m][n]$, we denote the set of $m\times n$ matrices over max-plus algebra. The binary operations on max-plus algebra can be extended to vectors and matrices in the natural way. Given $A,B\in \Rmax[m][n], C\in \Rmax[m][p],D\in \Rmax[p][n]$ and $\alpha\in \Rmax$,
\begin{align*}
[\alpha\otimes A](i,j)&=\alpha + A(i,j),\\
[A\oplus B](i,j)&=A(i,j) \oplus B(i,j),\\
[C\otimes D](i,j)&=\bigoplus_{k=1}^p C(i,k)\otimes D(k,j),
\end{align*}
for all $i=1,\ldots,m$ and $j=1,\ldots,n$.
For $A\in\Rmax[n][n]$ and $t\in \mathbb{N}$, $A^{\otimes t}$ denotes $A\otimes\ldots\otimes A$ ($t$ times). 
For $t=0$, $A^{\otimes 0}$ is an $n$-dimensional max-plus identity matrix whose  diagonal and non-diagonal elements are $0$ and $\varepsilon$, respectively.

For a set of vectors $U=\{u_1,\ldots,u_p\}$ in $\Rmax^n$, we use the same notation to denote a matrix whose columns are in $U$ i.e., $U(\cdot,i)=u_i$ for $1\leq i \leq p$. A vector $u\in \mathbb{R}^n$
is \textit{a max-plus linear combination} of $U$ if there exist some some scalars $\alpha_1,\ldots,\alpha_p\in \mathbb{R}$ such that $u=\alpha_1\otimes u_1\oplus \ldots \oplus \alpha_p \otimes u_p$
or equivalently there exists $w\in \mathbb{R}^p$ such that $U\otimes w=u$. The set of all max-plus linear combinations of $U$ is called the \textit{max-plus cone}
and is denoted by $\mathsf{cone}(U)$ \cite{Butkovic}. In this paper, we require max-plus cones to be subsets of $\mathbb{R}^n$. 

A dynamical system over the max-plus algebra is called a Max-Plus Linear (MPL) system, defined as 
\begin{equation}
\textbf{x}(k+1)=A\otimes \textbf{x}(k), ~~k=0,1,\ldots, 
\label{mpl}
\end{equation}
where $A\in \Rmax[n][n]$ is the system matrix, and vector $\textbf{x}(k)=[x_1(k)~\ldots~ x_n(k)]^\top$ encodes the state variables \cite{Baccelli}. 
The vector $\textbf{x}(k)$ is used to represent the time stamps associated to the discrete events at the $k$-th event counter. In this paper, we assume that $A$ is a \textit{regular} matrix i.e., there is at least one finite element in each row of $A$ \cite{Heidergott}.  Many applications of MPL models are found in dynamical systems, where modelling the time associated to discrete events is essential, such as in transportation networks \cite{Heidergott}, in scheduling \cite{Alirezaei} or manufacturing problems \cite{Aleksey}, or in biological systems \cite{Chris,Comet}. 

\begin{defn}[Orbit and Lasso \cite{Mufid2021}]\label{def:orbit}
Given an MPL system \eqref{mpl} with an initial vector $\textbf{x}(0)$, a sequence 
 $\textbf{x}(0)\textbf{x}(1)\ldots$ is called an orbit from $\textbf{x}(0)$ w.r.t. $A$. Furthermore, if there exist $\alpha\in\mathbb{R}$ and $k\geq l\geq 0$ such that $\textbf{x}(k+1+j)=\alpha\otimes \textbf{x}(l+j)$ for $j\geq 0$ then such sequence is called a $(k,l)$-lasso and $l$ is the \textit{loopback} bound. Figure \ref{lasso} depicts the illustration of a lasso.  
\end{defn}
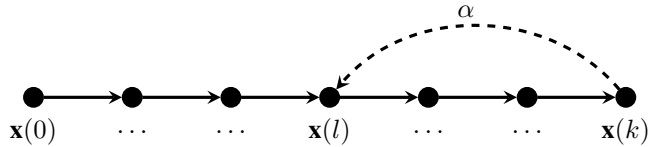
\begin{figure}[!ht]
\centering
    \begin{tikzpicture}[node distance=1.75cm and 1.75cm]
    \tikzstyle{state}=[circle,thick,draw=black,fill=black,minimum size=0.5mm,scale=0.75]
    \node[state] (1){};
    \node[state] (2) [right of = 1]{};
    \node[state] (3) [right of = 2]{};
    \node[state] (4) [right of = 3]{};
    \node[state] (5) [right of = 4]{} ;
    \node[state] (6) [right of = 5]{} ;
    \node[state] (7) [right of = 6]{} ;
    
    \draw[ ->,>=stealth,very thick] (1) to (2);
    \draw[ ->,>=stealth,very thick] (2) to (3);
    \draw[ ->,>=stealth,very thick] (3) to (4);
    \draw[ ->,>=stealth,very thick] (4) to (5);
    \draw[ ->,>=stealth,very thick] (5) to (6);
    \draw[ ->,>=stealth,very thick] (6) to (7);
    \draw[ ->,dashed,>=stealth,very thick] (7) to [bend right =50] (4);
    \node [below of = 1,yshift=1.3cm] {$\textbf{x}(0)$};
    \node [below of = 2,yshift=1.3cm] {$\ldots$};
    \node [below of = 3,yshift=1.3cm] {$\ldots$};
    \node [below of = 4,yshift=1.3cm] {$\textbf{x}(l)$};
    \node [below of = 5,yshift=1.3cm] {$\ldots$};
    \node [below of = 6,yshift=1.3cm] {$\ldots$};
    \node [below of = 7,yshift=1.3cm] {$\textbf{x}(k)$};
    \node [above of = 5,xshift=.5cm,yshift=-0.6cm] {$\alpha$};
\end{tikzpicture}
\caption{An illustration of a $(k,l)$-lasso. The dashed arrow represents the transition $\textbf{x}(k+1)=\alpha\otimes \textbf{x}(l)$.}
\label{lasso}
\end{figure}

Let us remark that an orbit represents the execution (or path) of the MPL system \eqref{mpl} originating from an initial state that is a vector. To avoid ambiguity, it should be noted that orbits and paths can be used interchangeably in this paper underlining the obvious fact that an MPL system is a special case of an STS, and an MPL orbit is equivalent to a path in the STS. The definition of lasso in \autoref{def:orbit}, instead, is slightly different from the usual one found in literature \cite{Biere1,Biere2}, which requires that the $l$-th and $(k+1)$-th states are exactly the same. In \autoref{sec-algorithms-model-checking} we will clarify the relationship between these two interpretations.

The notation $\mathsf{Orb}(A)=\{\textbf{x}(0)\textbf{x}(1)\ldots\mid \textbf{x}(0)\in \mathbb{R}^n\}$ represents the set of all orbits w.r.t. $A$. Likewise, $\mathsf{Orb}(A,X)=\{\textbf{x}(0)\textbf{x}(1)\ldots\mid \textbf{x}(0)\in X\}$ is the set of orbits w.r.t. $A$ starting from a set of initial conditions $X$. For the sake of simplicity we use the following notation to refer to an orbit:  $\pi=\textbf{x}(0)\textbf{x}(1)\ldots$. Given an orbit $\pi$ and $j\geq 0$, $\pi[j]=\textbf{x}(j)$ denotes the $j$-th vector of $\pi$ while $\pi[j..]$ is the $j$-th suffix of $\pi$, i.e., $\pi[j..]=\textbf{x}(j)\textbf{x}(j+1)\ldots$ . We say that orbit $\pi$ is \textit{similar} to orbit $\sigma $ iff there exists $\beta\in\mathbb{R}$ such that $\pi[0]=\beta\otimes \sigma[0]$ (which implies $\pi[j]=\beta\otimes \sigma[j]$ for $j\geq 0$).
\begin{defn}[Precedence Graph \cite{Baccelli}]
\label{precedence}
The precedence graph of $A\in\Rmax[n][n]$, denoted by $\mathcal{G}(A)$, is a weighted directed graph with nodes $1,\ldots,n$ and an edge from $j$ to $i$ with weight $A(i,j)$ for each $A(i,j)\neq \varepsilon$.
\end{defn}

A directed graph is strongly connected if, for any two different nodes $i,j$, there exists a path from $i$ to $j$. The weight of a path 
is equal to the sum of 
its edge-weights.
A circuit, namely a path that begins and ends at the same node, is called \textit{critical} if it has maximum average weight, which is the weight divided by the length of the path \cite{Baccelli}. The notion of precedence graph plays an important role for the analysis of MPL systems, such as for the eigenprolem \cite{Cuninghame,Subiono} and for the study of periodic behavior \cite{Nowak,Bart2,Molnarova}.


\begin{defn}[Irreducible Matrix \cite{Baccelli}]
A matrix $A\in \Rmax[n][n]$ is called \textit{irreducible} if $\mathcal{G}(A)$ is strongly connected.
\end{defn} 
For the rest of this paper, we define the irreducibility of an MPL system \eqref{mpl} as that of its state matrix $A\in \Rmax[n][n]$.
Each irreducible MPL system \eqref{mpl} admits a unique max-plus eigenvalue $\lambda\in\mathbb{R}$ and a corresponding max-plus eigenspace $E(A)=\{{x}\in\Rmax[n]\mid A\otimes {x}=\lambda\otimes {x}\}$.
The scalar $\lambda$ is equal to the average weight of critical circuits in $\mathcal{G}(A)$, while $E(A)$ can be computed from $A_\lambda^+=\bigoplus_{k=1}^n((-\lambda)\otimes A)^{\otimes k}$. 
A reducible MPL system may have multiple eigenvalues, where the maximum one equals to the average weight of critical circuits of $\mathcal{G}(A)$ \cite{Baccelli}. 
\begin{exmp}
\label{ex1}
Consider a two-dimensional MPL system 
\begin{equation}
\textbf{x}(k+1)=A\otimes \textbf{x}(k),~~
A =
\begin{bmatrix}
2 & 5 \\
3 & 3
\end{bmatrix},
\label{mpl_ex}
\end{equation}
that models an abstract railway network, as shown in \autoref{railway}, where $x_i(k)$ represents the time of the $k$-th departure at station $S_i$ for $i\in\{1,2\}$. The element $A(i,j)$ for $i\neq j$ corresponds to the time units needed to travel from station $S_j$ to $S_i$. The element $A(i,i)$ represents the delay required for the next departure of a train from station $S_i$. For instance, after a train departs from $S_1$ at the $m$-th time unit, then the next departure cannot be earlier than $(m+2)$ time units. 
The model's semantics are as follows: the max operation requires trains to synchronise their departure from stations, depending on all the previous departure times from connecting stations and the corresponding transfer times. 
\begin{figure}[!ht]
\centering
\includegraphics[scale=1]{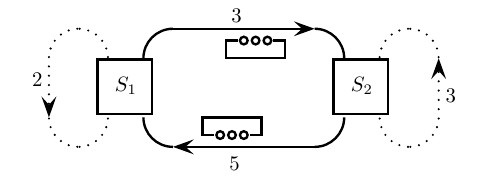}
\caption{A simple railway network represented by an MPL system in \eqref{mpl_ex}.}
\label{railway}
\end{figure}

The network in \autoref{railway} can be interpreted as the precedence graph $\mathcal{G}(A)$. As such, the critical circuit in $\mathcal{G}(A)$ is $S_1\rightarrow S_2\rightarrow S_1$. Furthermore, the max-plus eigenvalue is $\lambda=4$ with the eigenspace $E(A)=\{\textbf{x}\in\mathbb{R}^2\mid x_1-x_2=1\}$. Suppose the vector of initial departures is $\textbf{x}(0)=[0~0]^\top$, then its corresponding orbit is 
\[
\pi=
\begin{bmatrix}
0\\0
\end{bmatrix},
\begin{bmatrix}
5\\3
\end{bmatrix},
\begin{bmatrix}
8\\8
\end{bmatrix},
\begin{bmatrix}
13\\11
\end{bmatrix},\ldots
\]
Notice that, $\pi$ is a $(1,0)$-lasso. It is also straightforward to see that $\pi[0..]$ is similar to $\pi[2..]$.
\end{exmp}
\section{Verification Problems for MPL systems}
\label{sec-problem-definition}

This section presents two problems over MPL systems. The first one relates to the periodic behavior of MPL systems, and in particular it deals with detecting the length of transients and the period of cyclic behaviors. The second problem concerns the verification of Time-Difference Linear Temporal Logic. This extends the reachability analysis problem for MPL systems discussed in~\cite{DiekyBackward,DiekyForward}.
\subsection{Transient and Cyclicity}


The transient and cyclicity of MPL systems \eqref{mpl} is related to the sequence of the powers of its state matrix $A$, namely $A^{\otimes{k}}$ for $k\geq 0$.
\begin{prop}[Transient \cite{Baccelli,Heidergott}] 
\label{transient_bound}
For an irreducible matrix $A\in\Rmax^{n\times n}$ and its max-plus eigenvalue $\lambda\in\mathbb{R}$, there exist integers $l,c\geq 0$, such that $A^{\otimes(k+c)}=(\lambda\times c)\otimes A^{\otimes k}$ for all $k\geq l$. The smallest such $l$ and $c$ are called the \textit{transient} and the \textit{cyclicity} of $A$, respectively. 
\label{trans}
\end{prop}
For the rest of this paper, the transient and cyclicity of $A$ are defined by $\mathsf{tr}(A)$ and $\mathsf{cyc}(A)$, respectively. The cyclicity $\mathsf{cyc}(A)$ is related to critical circuits in the precedence graph $\mathcal{G}(A)$ \cite[Definition 3.94]{Baccelli} and its computation is relatively straightforward. On the other hand, the transient $\mathsf{tr}(A)$ is harder to compute since it is unrelated to the dimension of $A$.  The transient of $A\in \Rmax[n][n]$ can be really large even for a small $n$. Upper bounds for the transient have been discussed in \cite{Charron,Merlet,Nowak,Gerardo}.  

Due to \autoref{trans}, the orbits of an irreducible MPL system induce an \textit{eventually periodic} behaviour with a rate $\lambda$: for each initial vector $\textbf{x}(0)\in \mathbb{R}^n$ we have $\textbf{x}(k+\mathsf{cyc}(A))=(\lambda \times \mathsf{cyc}(A))\otimes \textbf{x}(k)$ for all $k\geq \mathsf{tr}(A)$. One can say that the $(k+\mathsf{cyc}(A))$-th state of an irreducible MPL system is equal to the ``shifting'' of the $k$-th state w.r.t a scalar $\lambda\times\mathsf{cyc}(A)$, namely $x_i(k+\mathsf{cyc}(A))=x_i(k)+\lambda\times c$ for $i\in \{1,\ldots,n\}$.  A similar condition may be found on 
reducible MPL systems: we denote the corresponding transient and cyclicity to be global, as per \autoref{transient_bound}. The local transient and cyclicity for a specific initial vector $\textbf{x}(0)\in\mathbb{R}^n$ and for a set of initial vectors $X\subseteq \mathbb{R}^n$ have been studied in \cite{Mufid2020} and are defined as follows.  
\begin{defn}[\cite{Mufid2020}]
Given $A\in\Rmax[n][n]$ with eigenvalue $\lambda$ and an initial vector $\textbf{x}(0)\in \mathbb{R}^n$, the local transient and cyclicity of $A$ w.r.t. $\textbf{x}(0)$ are respectively the smallest $l,c\in \mathbb{N}_0$ such that
$\textbf{x}(j+c)=\lambda c\otimes \textbf{x}(j),\forall j\geq l$.
We denote those scalars as $\mathsf{tr}(A,\textbf{x}(0))$ and $\mathsf{cyc}(A,\textbf{x}(0))$, respectively. Furthermore, for $X\subseteq \mathbb{R}^n$, $\mathsf{tr}(A,X)=\texttt{sup}\{\mathsf{tr}(A,\textbf{x}(0))\mid \textbf{x}(0)\in X\}$ and $\mathsf{cyc}(A,X)=\texttt{lcm}\{\mathsf{cyc}(A,\textbf{x}(0))\mid \textbf{x}(0)\in X\}$, where $\texttt{lcm}$ is the least common multiple.  
\label{trans_set}
\end{defn}

Following \cite{Mufid2020}, any MPL system \eqref{mpl} can be categorised into three classes: 1) \textit{never periodic} if  $\mathsf{tr}(A,\textbf{x}(0))$ does not exist for all $\textbf{x}(0)\in \mathbb{R}^n$; 2) \textit{boundedly periodic} if $\mathsf{tr}(A,\textbf{x}(0))$ exists for all $\textbf{x}(0)\in \mathbb{R}^n$ and $\mathsf{tr}(A)$ exists; and 3) \textit{unboundedly periodic} if $\mathsf{tr}(A,\textbf{x}(0))$ exists for all $\textbf{x}(0)\in \mathbb{R}^n$ but $\mathsf{tr}(A)$ does not. We call \eqref{mpl} \textit{periodic} if it is either \textit{unboundedly periodic} or \textit{boundedly periodic}. 

The periodic behavior of MPL systems is closely related to the cycle-time vector, which corresponds to the asymptotic behavior of $\textbf{x}(k)/k$ as $k\to +\infty$.
\begin{defn}[Cycle-Time Vector \cite{Heidergott}]
\label{cycle_time}
Consider a regular MPL system \eqref{mpl}, and assume that for all $j\in \{1,\ldots,n\}$ the quantity $\eta_j$, defined by 
\[\displaystyle \eta_j=\lim_{k\rightarrow +\infty} (x_j(k)/k),\]
exists. Then the vector $\chi=[\eta_1~\ldots~\eta_n]^\top$ is called the \textit{cycle-time vector} of the given sequence $\textbf{x}(k)$ with respect to $A$. 
\end{defn}

It has been shown in \cite[Theorem 3.11]{Heidergott} that if the cycle-time vector of $A$ exists for at least one initial vector,  then it exists for any initial vector. The cycle-time vector can be generated using the procedure in \cite[Algorithm 31]{Fahim}. 

\autoref{periodic_everywhere} shows that the periodic behavior of an MPL system is indeed related to the eigenspace and cycle-time vector of its corresponding state matrix. Furthermore, Propositions \ref{lasso_trans} and \ref{all_orbits} indicate the equivalent relation between periodic behavior and lasso orbits. 
\begin{prop}[\cite{AbateCMM20}]
\label{periodic_everywhere}
Suppose we have a regular matrix $A\in \Rmax[n][n]$ with a maximum eigenvalue $\lambda$ and cycle-time vector $\chi$. The following statements are equivalent: 
\begin{itemize}
    \item[$a$.] The underlying MPL system \eqref{mpl} is \textit{periodic}.
    \item[$b$.] The cycle-time vector is $\chi=[\lambda~\ldots~\lambda]^\top\in \mathbb{R}^n$.
    \item[$c$.] The eigenspace $E(A)$ is not empty. 
\end{itemize}
\end{prop}

\begin{prop}[\cite{Mufid2021}]
\label{lasso_trans}
An orbit $\pi$ is a lasso iff $\pi[0]$ admits local transient and cyclicity. 
\end{prop}
\begin{prop}[\cite{Mufid2021}]
\label{all_orbits}
An MPL system, as in \eqref{mpl}, is periodic iff all orbits $\pi\in\mathsf{Orb}(A)$ are lassos. 
\end{prop}

\begin{exmp}
\label{ex:transient}
From the preceding MPL system and initial vector in \autoref{ex1}, the corresponding orbit $\pi$ is periodic with transient $\mathsf{tr}(A)=0$ and cyclicity $\mathsf{cyc}(A)=2$.
\end{exmp}
In general, as one can see from \autoref{ex1} and \autoref{ex:transient}, if an initial vector $\textbf{x}(0)$ is periodic with $\mathsf{tr}(A,\textbf{x}(0))=l$ and $\mathsf{cyc}(A,\textbf{x}(0))=c$, then its corresponding orbit $\pi=\textbf{x}(0)\textbf{x}(1)\ldots$ is a $(k,l)$-lasso with $k=l+c-1$. Furthermore, the orbit $\pi[(l+c)..]$ is indeed similar to $\pi[k..]$.

\ifthenelse{\boolean{papertwocol}}{%
\begin{algorithm}[H]
	\footnotesize
	\captionsetup{format=hang,width=\linewidth}
    \caption{Computation of cyclicity and transient of $A$ w.r.t. $\mathsf{cone}(U)$}
    \label{comp_cyc_trans}
    \centering
    \begin{algorithmic}[1]
\Function{TransCone}{$A,U,N$}
\State $\textbf{M}\gets\Call{EmptyVector}$ \Comment{empty vector used to store}
\State $\textbf{M}.\mathsf{push}\_\mathsf{back}(U)$ \CommentX{$A^{0}\otimes U,A^{1}\otimes U,\ldots$\hspace{5.0ex}}
\State $it\gets 0$\Comment{number of iterations}
\State $\chi\gets \Call{CycleTimeVector}{A}$ \Comment{computing cycle-time vector}
\If{elements of $\chi$ are all equal}
\State $\lambda\gets \chi[0]$ \Comment{$\lambda$ is the eigenvalue of $A$}
\While{($it\leq N)$}
\State $\textbf{M}.\mathsf{push}\_\mathsf{back}(A\otimes \textbf{M}[it])$
\State $it\gets it+1$
\For{$1\leq m<it $}
\If{$(\textbf{M}[it]=(\lambda\times m)\otimes \textbf{M}[it-m])$}
\State{\Return{$\tuple{it-m, m}$}}
\EndIf
\EndFor
\EndWhile
\If{$(it> N)$}
\State \textbf{print} ``terminated after reaching maximum bound''
\EndIf
\Else
\State \textbf{print} ``the transient does not exist''
\EndIf
\EndFunction
    \end{algorithmic}
\end{algorithm}
The classical method to compute transient and cyclicity for $A\in \Rmax[n][n]$ is by employing matrix-multiplication technique as in \autoref{comp_cyc_trans}. It should be noticed that the algorithm can only be applied on a max-plus cone represented by a matrix $U$. The algorithm starts by computing the cycle-time vector $\chi$ for $A$. If the entries of $\chi$ are not all the same, then the transient for $\mathsf{cone}(U)$ does not exist. On the other hand, if elements of $\chi$ are all equal (and  correspond to the eigenvalue of $A$, as per \autoref{periodic_everywhere}), then we compute $U,A\otimes U, A^{\otimes 2}\otimes U,\ldots$. This process is repeated until we find $it>m\geq 1$ such that $A^{\otimes it}\otimes U=A^{\otimes (it-m)}\otimes U$. The values of $it-m$ and $m$ correspond to transient and cyclicity, respectively. \autoref{comp_cyc_trans} can also be used to compute the local transient and cyclicity for a vector: that is, when $U$ has only one column.
}
{
The classical method to compute transient and cyclicity for $A\in \Rmax[n][n]$ is by employing a matrix-multiplication technique, as in \autoref{comp_cyc_trans}. It should be noticed that the algorithm can only be applied on a max-plus cone represented by a matrix $U$. The algorithm starts by computing the cycle-time vector $\chi$ for $A$. If the entries of $\chi$ are not all the same, then the transient for $\mathsf{cone}(U)$ does not exist. On the other hand, if elements of $\chi$ are all equal (and  correspond to the eigenvalue of $A$, as per \autoref{periodic_everywhere}), then we compute $U,A\otimes U, A^{\otimes 2}\otimes U,\ldots$. This process is repeated until we find $it>m\geq 1$ such that $A^{\otimes it}\otimes U=A^{\otimes (it-m)}\otimes U$. The values of $it-m$ and $m$ correspond to transient and cyclicity, respectively. \autoref{comp_cyc_trans} can also be used to compute the local transient and cyclicity for a vector, that is, when $U$ has only one column.
\begin{algorithm}[H]
	\footnotesize
	\captionsetup{format=hang,width=\linewidth}
    \caption{Computation of cyclicity and transient of $A$ w.r.t. $\mathsf{cone}(U)$}
    \label{comp_cyc_trans}
    \centering
    \begin{algorithmic}[1]
\Function{TransCone}{$A,U,N$}
\State $\textbf{M}\gets\Call{EmptyVector}$ \Comment{empty vector used to store $A^{0}\otimes U,A^{1}\otimes U,\ldots$}
\State $\textbf{M}.\mathsf{push}\_\mathsf{back}(U)$
\State $it\gets 0$\Comment{number of iterations}
\State $\chi\gets \Call{CycleTimeVector}{A}$ \Comment{computing cycle-time vector}
\If{elements of $\chi$ are all equal}
\State $\lambda\gets \chi[0]$ \Comment{$\lambda$ is the eigenvalue of $A$}
\While{($it\leq N)$}
\State $\textbf{M}.\mathsf{push}\_\mathsf{back}(A\otimes \textbf{M}[it])$
\State $it\gets it+1$
\For{$1\leq m<it $}
\If{$(\textbf{M}[it]=(\lambda\times m)\otimes \textbf{M}[it-m])$}
\State{\Return{$\tuple{it-m, m}$}}
\EndIf
\EndFor
\EndWhile
\If{$(it> N)$}
\State \textbf{print} ``terminated after reaching maximum bound''
\EndIf
\Else
\State \textbf{print} ``the transient does not exist''
\EndIf
\EndFunction
    \end{algorithmic}
\end{algorithm}
}

It is straightforward to see that computing the transient is decidable  for \textit{boundedly periodic} MPL systems and undecidable for \textit{unboundedly periodic} ones. For this reason, we assign a maximum bound $N$ as a general termination condition. A similar setting is also used in the TDLTL model checking algorithms described in Sections \ref{sec-algorithms-model-checking} and \ref{sec-algorithms-smt}.

\subsection{TDLTL Model Checking}

In this section we first introduce a temporal logic, called Time-Difference Linear Temporal Logic (TDLTL), which we devise for expressing verification queries over MPL systems, then we formalize the TDLTL model-checking problem for MPL systems. Intuitively, TDLTL is obtained from standard LTL, as defined in Section~\ref{sec-preliminaries}, by considering a special class of atoms, called~\emph{time-difference propositions}, where it is possible to compare the values of state variables at different time points. Moreover, the model checking problem is defined analogously to the case of STSs with the difference that MPL orbits are used instead of STS paths.

\begin{defn}
\label{TD_prop}
A time-difference (TD) proposition over a vector of variables $\vec{\texttt{x}} = \langle \texttt{x}_1, \cdots, \texttt{x}_n \rangle$ is an atomic formula $ p= \texttt{x}^{(k)}_{i}-\texttt{x}^{(l)}_{j}\sim \alpha$, where $i,j\in \{1,\ldots,n\}, k,l \in \mathbb{N}, \alpha\in \mathbb{R}$ and $\sim\in\{>,\geq \}$.
We call $p$ ``initial'' if $k=l=0$: for the sake of simplicity, we write $\texttt{x}_i-\texttt{x}_j\sim \alpha$ instead. 
For $m\geq 0$, we write $p^{(m)}=\texttt{x}^{(k+m)}_{i}-\texttt{x}^{(l+m)}_{j}\sim \alpha$.
\end{defn}

\begin{defn}
\label{TD_formula}
A TD formula for a vector of variables $\vec{x}$ is defined according to the following grammar
$$f::=\top~\mid ~ p~\mid ~\neg f~ \mid ~f_1\wedge f_2~ \mid ~f_1\vee f_2,$$
where $p= \texttt{x}^{(k)}_{i}-\texttt{x}^{(l)}_{j}\sim \alpha$ is a TD proposition over $\vec{\texttt{x}}$. We call a TD formula $f$ initial if all TD propositions appearing in $f$ are initial ones. 
\end{defn}

The satisfaction relation of TD formula $f$ is defined on an orbit $\pi$ as follows:  
\begin{equation}
\label{TD_formula_semantic}
\left.
\begin{array}{ll}
\pi \models \top,&\\
\pi\models \texttt{x}^{(k)}_{i}-\texttt{x}^{(l)}_{j}\sim \alpha& ~\text{iff}~ \pi[k]_i \sim \pi[l]_j+\alpha,\\
\pi\models  \neg f_1& ~\text{iff}~ \pi\not\models f_1,\\
\pi \models f_1\wedge f_2& ~\text{iff}~\pi\models f_1\wedge \pi\models f_2,\\
\pi \models f_1\vee f_2& ~\text{iff}~\pi\models f_1\vee \pi\models f_2.\\
\end{array}\right\} 
\end{equation}

Intuitively and unsurprisingly, we use the standard definition for Boolean connectives. Time-difference propositions, instead, are interpreted on the orbit $\pi$ by using the values assigned to the system variables at different times.

In the case of MPL systems, an orbit is uniquely determined by its initial vector $\textbf{x}(0)=\pi[0]$; hence, one can write
\begin{equation}
\label{state_TD}
    \pi[0] \models f ~\text{iff}~\pi\models f. 
\end{equation}
Finally, given an MPL system \eqref{mpl} with a set of initial conditions $X\subseteq \mathbb{R}^n$ and a TD formula $f$, we write $(A,X) \models f$ if all orbits of $A$ starting from $X$ satisfy $f$ i.e., $\forall \pi \in \mathsf{Orb}(A,X).~ \pi \models f$.

We can now define the syntax and semantics of Time-Difference LTL (TDLTL) specifications. Essentially, TDLTL is a logic derived from LTL where atoms are TD-propositions. More formally, the following grammar defines the syntax of TDLTL formulae:
\begin{equation}
\varphi:=\top~|~p~|~\varphi_1 \wedge \varphi_2~|~\neg \varphi~|
~\ltlX \varphi~|~\varphi_1 ~\ltlU~\varphi_2
\label{TDLTL_syntact}
\end{equation}
where $p$ is an initial TD proposition. 
In this paper, without loss of generality, we assume that the TDLTL formulae are in positive normal form.
The benefit of the use of PNF formulae is that the negations ``can be pushed'' into the propositions. Notice that, the negation of a TD proposition in \autoref{TD_prop} yields another TD proposition\footnote{For example, the negation of $x_i-x_j>\alpha$ is $x_j-x_i\geq -\alpha$.}.


\ifthenelse{\boolean{papertwocol}}{
The semantics of TDLTL formulae \eqref{TDLTL_syntact} are defined over an orbit $\pi$. For a TD proposition $p$, $\pi \models p$ iff $\pi[0]\models p$ and $\pi\models \neg p$ iff $\pi \not\models p$. The semantics for TDLTL formulae with Boolean and temporal operators are defined similarly as in \eqref{LTL_semantic}.
}{%
The semantics of TDLTL formulae \eqref{TDLTL_syntact} is defined as on an orbit $\pi$ as follows:
\begin{equation}
\left.
\label{TDLTL_semantic}
\begin{array}{ll}
\pi\models p& \text{iff}~ \pi[0]\models p,\\
\pi\models  \neg p& \text{iff}~ \pi\not\models p,\\
\pi\models \varphi_1\wedge \varphi_2& \text{iff}~\pi\models \varphi_1\wedge \pi\models\varphi_2,\\
\pi\models \varphi_1\vee \varphi_2& \text{iff}~\pi\models \varphi_1\vee \pi\models\varphi_2,\\
\pi\models \ltlX \varphi& \text{iff}~\pi[1..]\models \varphi,\\
\pi\models \varphi_1 ~\ltlU~\varphi_2& \text{iff}~\exists j\geq 0.~\pi[j..]\models \varphi_2~\text{and}~\forall 0\leq i< j.~\pi[i..]\models \varphi_1,\\
\pi\models \varphi_1 ~\ltlR~\varphi_2& \text{iff}~\forall j\geq 0.~\pi[j..]\models \varphi_2~\text{or}~\exists i\geq 0.~(\pi[i..]\models \varphi_1\wedge \forall h\leq i.~ \pi[h..]\models \varphi_2 ),\\
 \pi\models  \ltlF \varphi& \text{iff}~\exists j\geq 0.~\pi[j..]\models \varphi,\\
 \pi\models  \ltlG \varphi&\text{iff}~\forall j\geq 0.~\pi[j..]\models \varphi.
\end{array}\!\!\!\right\}
\end{equation}
}
Similar to \eqref{state_TD}, the semantics of TDLTL formulae over initial vectors in the case of an MPL system is as follows: $\pi[0] \models \varphi ~\text{iff}~\pi\models \varphi$. Given an MPL system \eqref{mpl} with a set of initial conditions $X\subseteq\mathbb{R}^n$ and a TDLTL formula $\varphi$, we write
\begin{equation}
\label{BMC_problem}
    (A,X) \models \varphi~\text{iff}~\forall \pi\in \mathsf{Orb}(A,X).~ \pi \models\varphi. 
\end{equation} 

\section{Algorithms based on Reduction to Infinite-State Transition Systems}
\label{sec-algorithms-model-checking}

In this section, we show how the problems of finding the transient and cyclicity of an MPL system and of checking TDLTL properties can be reduced to the problem of LTL model-checking of infinite-state STSs. In particular, we start by showing how we can check if a given transient $k_0$ and cyclicity $c$ are sufficient for a given set of initial states. Then, we will show how to encode TDLTL model-checking into plain LTL model-checking of infinite-state transition systems.

Before delving into the two problems, we give a basic encoding of the MPL dynamics as a STS. This encoding is common to all the model-checking reduction approaches defined below and targets a STS systems with \qflra formulae used for the transition relation and the initial states. We highlight that we assume that the initial states can be encoded as a single \qflra formula $\iota$, meaning that every model (satisfying value) of $\iota$ is a possible initial state for the system. We recall that to define an STS, we use a set of variables and in the transition relation we use the notation $x'$ to indicate the value of variable $x$ in the next state (see \autoref{sec-preliminaries} for details). 

\begin{defn}[STS Encoding of MPL Systems]\label{def:sts-basic}
Given an MPL system \eqref{mpl}
and an initial state constraint $\iota$ in \qflra, the STS encoding the traces of the MPL system is $\langle V, \iota, T \rangle$, where 
\begin{itemize}
    \item $V \dot{=} \{x_1, \ldots, x_{n}\}$ where all the variables are of real type;
    \item $T \dot{=} (\bigwedge_{i=1}^n \bigwedge_{j=1}^n x'_i \ge x_j + A_{i,j}) \wedge (\bigwedge_{i=1}^n \bigvee_{j=1}^n x'_i = x_j + A_{i,j})$.
\end{itemize}
\end{defn}

Intuitively, we construct an STS that has the same variables as the MPL system and that is such that every orbit of the MPL corresponds to a path of the STS. The encoding of the transition relation is the most interesting part of the construction: for example, if we want to enforce the equality $x'_i = \max\{\phi_1, \ldots, \phi_n\}$, we impose two transition constraints: we enforce $x'_i$ to be greater of equal to all the elements $\{\phi_1, \ldots, \phi_n\}$ and we additionally require that $x'_i$ is equal to at least one element in $\{\phi_1, \ldots, \phi_n\}$: the only valid solution to such a constraint combination is the required maximum value. This encoding allows for an easier and more efficient handling of the maximization part of the MPL semantics, which represents a non-linearity in the system evolution.

The STS encoding in Definition \ref{def:sts-basic} directly captures the MPL system dynamics but can be problematic for actual model-checking algorithms because the concrete values assigned to the $x_i$ variables often diverge, while many model-checking algorithms look for cyclic behaviors (i.e. paths such that after a finite number of steps a loop of finite length is reached). We thus re-formulate our STS encoding transforming MPL loops into real loops 
by non-deterministically guessing a value $\lambda$ that witnesses the cycle, as detailed in the following. 

\begin{defn}[Looping STS Encoding of MPL Systems]\label{def:sts-looping}
Given an MPL system \eqref{mpl}
and an initial state constraint $\iota$ in \qflra, the looping STS encoding the traces of the MPL system is $\langle V, \iota, T \rangle$, where 
\begin{itemize}
    \item $V \dot{=} \{x_1, \cdots, x_{n}\} \cup \{ \lambda \}$ where all the variables are of real type;
    \item $T \dot{=} (\bigwedge_{i=1}^n \bigwedge_{j=1}^n x'_i \ge x_j + A_{i,j} - \lambda) \wedge (\bigwedge_{i=1}^n \bigvee_{j=1}^n x'_i = x_j + A_{i,j} - \lambda) \wedge (\lambda' = \lambda)$. 
\end{itemize}
\end{defn}
The variable $\lambda$ in \autoref{def:sts-looping} represents the eigenvalue of $A$ which, for irreducible MPL systems, is unique. Intuitively, the value of $\lambda$ must be guessed in the first step and then it should be constant at the next step (the constraint $\lambda' = \lambda$ forces the next value of $\lambda$ to be equal to the current value): $\lambda$ is a so-called \emph{frozen} variable. 
Any looping behavior of the STS system corresponds to an MPL looping behavior. Also, if we know the value of $\lambda$ (by for example computing the eigenvalue of an irreducible matrix) we can substitute that value in the encoding.

\begin{prop}
Given an MPL system \eqref{mpl} and a path $s_1, s_2, \cdots$ in its Looping STS Encoding, the sequence 
\ifthenelse{\boolean{papertwocol}}{
\begin{align*}
&[s_1(x_1),\ldots,s_1(x_n)]^\top, [s_2(x_1)\!+\!\bar{\lambda},\ldots,s_2(x_n)\!+\!\bar{\lambda}]^\top,\\
&\ldots, [s_i(x_1)\!+\!(i\!-\!1)\!\times\! \bar{\lambda},\ldots,s_i(x_n)\!+\!(i\!-\!1)\!\times\! \bar{\lambda}]^\top, \ldots
\end{align*}
}{
\[
[s_1(x_1),\ldots,s_1(x_n)]^\top, [s_2(x_1)+\bar{\lambda},\ldots,s_2(x_n)+\bar{\lambda}]^\top, \ldots, [s_i(x_1)+(i-1)\times \bar{\lambda},\ldots,s_i(x_n)+(i-1)\times \bar{\lambda}]^\top, \ldots
\]
}
where $\bar{\lambda} = s_1(\lambda)$, is a valid orbit of $M$.
\end{prop}

It easy to see that if a path in the Looping STS Encoding has a loop such that $s_l = s_k$ with $k<l$, then the MPL system has a $(k,l)$-lasso orbit, as per Definition~\ref{def:orbit}.

\subsection{Algorithms for Transient and Cyclicity Checking}

We now move to the decision problem of checking whether a given MPL system \eqref{mpl} has cyclicity $c$ and transient $k_0$. In order to check this property we can build on the STS encoding discussed above, but we need additional variables to keep track of the history of the system. In fact, we want to perform the check in pure-LTL and we essentially need to impose the requirement that any path of the STS encoding is such that the difference between the variables at step $k_0+c$ and the variables at step $k_0$ is a vector made up of the constant $\lambda$; in symbols: for every path $s_0,s_1, \cdots, s_{k_0+c}$, $s_{k_0+c}-s_{k_0} = [\lambda, \cdots, \lambda]^\top$. Since plain LTL, differently from TDLTL, does not admit atomic predicates across different times, we need to introduce ``history variables'' that record the values of the system variables at step $k_0$, so that we can compare them with the system variables at time $k_0+c$.

\begin{defn}[STS Encoding for Transient and Cyclicity Checking]
\label{def:sts-encoding-trans-cyc}
Given an MPL system \eqref{mpl} with an initial state constraint $\iota$ in \qflra, the STS encoding of the transient and cyclicity checking problem (indicated STS-TCC(A, $\iota$)), is an STS $\langle V', I, T' \rangle$, such that:
\begin{itemize}
    \item $V' \dot{=} V \cup \{h_1, \cdots, h_{n}\} \cup \{f\}$ where all the $h_i$ variables are of real type and $f$ is a Boolean variable;
    \item $I \dot{=} \iota \wedge (\neg f) \wedge (\bigwedge_{i=1}^n x_i = h_i)$;
    \item $T' \dot{=} T \wedge (f \rightarrow f') \wedge (f' \bigwedge_{i=1}^n h_i = h'_i) \wedge (\neg freeze \rightarrow \bigwedge_{i=1}^n x_i = h_i)$.
\end{itemize}
Here $\langle V, \iota, T \rangle$ is the STS looping encoding of the MPL system.
\end{defn}

The encoding of the transient and cyclicity checking is such that when the variable $f$ is assigned to true (this is possible at any time, as it starts from being false and it can be non-deterministically assigned to true anytime), it is forced to remain true for the rest of the path. The history variables $h_i$ are kept aligned with the system variables $x_i$ while $f$ is false, and are instead kept unchanged from the moment $f$ is set to true.

With this encoding, it is easy to write an LTL property that checks if a given transient $k_0$ and cyclicity $c$ are large enough for every orbit starting from a state satisfying an initial condition in $\iota$. 

\begin{prop}\label{prop:ltl_cyclicity}
Given an MPL system with \eqref{mpl} with an initial state constraint $\iota$ in \qflra, the system has a transient $k_0$ and a cyclicity $c$ if $k_0$ and $c$ are the minimal integers for which the following LTL property (indicated as TCC-Prop($k_0$, $c$)) is valid for the STS Encoding for Transient and Cyclicity Checking:
\ifthenelse{\boolean{papertwocol}}{%
\vskip -20pt
$$
\ltlX^{k_0} ((\neg f \wedge (\ltlX f)) \!\rightarrow\! \ltlX^{c} (\bigwedge_{i=1}^{n-1} (x_i -  h_i) = (x_{i+1} - h_{i+1}) )). 
$$
}{
$$
\mbox{TCC-Prop($k_0$, $c$)} \, \dot{=} \, \ltlX^{k_0} ((\neg f \wedge (\ltlX f)) \rightarrow \ltlX^{c} (\bigwedge_{i=1}^{n-1} (x_i -  h_i) = (x_{i+1} - h_{i+1}) )). 
$$
}
\end{prop}

The intuition behind Proposition \ref{prop:ltl_cyclicity} is that we focus on the paths to set $f$ to true at step $k_0$ and then we check at step $k_0+c$ whether the difference between the vectors $\vec{x}$ and $\vec{h}$ is a vector of the same constants. To do so, we use an implication and a nesting of $(k_0+c)$ LTL operators $\ltlX$ (next).

\begin{figure}[!ht]
    \centering
    \ifthenelse{\boolean{papertwocol}}{
    \resizebox{\linewidth}{!}{\footnotesize\begin{tikzpicture}[node distance = 2cm, auto,scale=1]
\node [par,text width=9em] (mpls) {
$\begin{array}{l}
A\in \mathbb{R}_{\max}^{n\times n}\\
\iota \subseteq \mathbb{R}^n
\end{array}$
};    
\node [block,text width=17em,right of = mpls,xshift=1.8cm,yshift=-1.5cm] (mat) { 
$\begin{array}{l}
M \gets \mbox{STS-TCC(A, $\iota$)}\\
\varphi \gets \mbox{TCC-Prop($k_0$, $c$)}
\end{array}$
};
\node [block,text width=12.5em,right of = mat,xshift=3.5cm] (SMT1) {$\mathsf{ModelChecking}(M, \varphi)$};
\node [par,text width=10em,above of = SMT1,yshift=-0.6cm] (out1) {Transient is $k_0$, cyclicity is $c$};    
\node [block,text width=12em,below of = SMT1] (model) {
$\begin{array}{l}
\pi \gets \mathsf{get}\_\mathsf{couterexample}()\\
k_0^\prime \gets \mathsf{tr}(A,\pi[0])\\
c^\prime \gets \mathsf{cyc}(A,\pi[0])
\end{array}$
};    
\node [block,left of = model,text width=5em,xshift=-7.3cm] (tr) { };    
\node [left of = tr] (emp) {};    
\path [line] (mpls) |- ($(mat.west)+(0,0.1)$); 
\path [line,dashed] (emp) -- node [xshift=0cm,yshift=-0.5cm] {\scriptsize 
$\begin{array}{l}
k_0\gets 0\vspace*{1ex}\\
c\gets 1
\end{array}$
}(tr); 
\path [line] (mat) -- (SMT1); 
\path [line] (SMT1) -- node [anchor=west] {\scriptsize $\mathsf{Valid}$}(out1); 
\path [line] (SMT1) -- node [anchor=west] {\scriptsize $ \mathsf{Invalid}$}(model); 
\path [line] (model) -- node [xshift=0.cm,yshift=0.45cm] {\scriptsize 
$\begin{array}{l}
k_0\gets \max\{k_0,k_0^\prime\}\vspace*{1ex}\\
c\gets \mathtt{lcm}\{c,c^\prime\}
\end{array}$
}(tr); 
\path [line] (tr) |- ($(mat.west)+(0,-0.1)$); 
\end{tikzpicture}}
    }{
    \resizebox{.6\linewidth}{!}{\footnotesize\begin{tikzpicture}[node distance = 2cm, auto,scale=1]
\node [par,text width=9em] (mpls) {
$\begin{array}{l}
A\in \mathbb{R}_{\max}^{n\times n}\\
\iota \subseteq \mathbb{R}^n
\end{array}$
};    
\node [block,text width=17em,right of = mpls,xshift=1.8cm,yshift=-1.5cm] (mat) { 
$\begin{array}{l}
M \gets \mbox{STS-TCC(A, $\iota$)}\\
\varphi \gets \mbox{TCC-Prop($k_0$, $c$)}
\end{array}$
};
\node [block,text width=12.5em,right of = mat,xshift=3.5cm] (SMT1) {$\mathsf{ModelChecking}(M, \varphi)$};
\node [par,text width=10em,above of = SMT1,yshift=-0.6cm] (out1) {Transient is $k_0$, cyclicity is $c$};    
\node [block,text width=12em,below of = SMT1] (model) {
$\begin{array}{l}
\pi \gets \mathsf{get}\_\mathsf{couterexample}()\\
k_0^\prime \gets \mathsf{tr}(A,\pi[0])\\
c^\prime \gets \mathsf{cyc}(A,\pi[0])
\end{array}$
};    
\node [block,left of = model,text width=5em,xshift=-7.3cm] (tr) { };    
\node [left of = tr] (emp) {};    
\path [line] (mpls) |- ($(mat.west)+(0,0.1)$); 
\path [line,dashed] (emp) -- node [xshift=0cm,yshift=-0.5cm] {\scriptsize 
$\begin{array}{l}
k_0\gets 0\vspace*{1ex}\\
c\gets 1
\end{array}$
}(tr); 
\path [line] (mat) -- (SMT1); 
\path [line] (SMT1) -- node [anchor=west] {\scriptsize $\mathsf{Valid}$}(out1); 
\path [line] (SMT1) -- node [anchor=west] {\scriptsize $ \mathsf{Invalid}$}(model); 
\path [line] (model) -- node [xshift=0.cm,yshift=0.45cm] {\scriptsize 
$\begin{array}{l}
k_0\gets \max\{k_0,k_0^\prime\}\vspace*{1ex}\\
c\gets \mathtt{lcm}\{c,c^\prime\}
\end{array}$
}(tr); 
\path [line] (tr) |- ($(mat.west)+(0,-0.1)$); 
\end{tikzpicture}}
    }    
    \captionsetup{format=hang,width=\linewidth}
\caption{Transient and cyclicity computation using STS model checking. The function $\mathsf{ModelCheck}$ represents the call to the LTL model-checker.}
\label{fig:MC-TCC}
\end{figure}
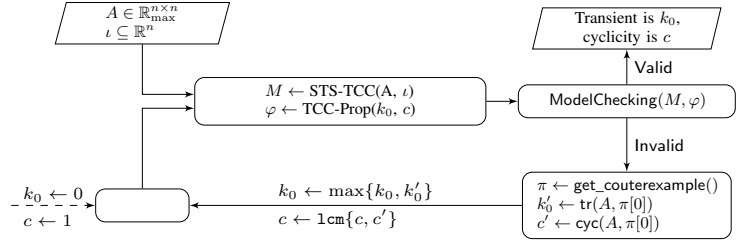

Note that the LTL property of Proposition \ref{prop:ltl_cyclicity} only proves that the given $k_0$ and $c$ are sufficient, not that they are minimal. For this reason one must explore the space of assignments to $k_0$ and $c$. The algorithm in \autoref{fig:MC-TCC} depicts a possible approach for this problem. The algorithm starts with an initial guess $k_0=0$ and $c=1$ and checks if this guess is correct by means of \autoref{prop:ltl_cyclicity}. If the model checker validates the property then $k_0=0$ and $c=1$ are indeed the transient and cyclicity for the MPL system as they are the minimum possible values. Otherwise, we extract the initial state from the counterexample trace of the model-checker and we extract the transient and cyclicity from this initial state. It must be the case that either the observed $k_0'>k_0$ or $c'$ is not a divisor of $c$ (otherwise $\pi$ would not have been a counterexample). At this point, a new guess of $k_0$ that takes value $\max\{k_0,k_0'\}$ and $c$ that takes value $\mathtt{lcm}\{c,c^\prime\}$ is made and the cycle repeats. The algorithm is guaranteed to find the minimum value for $k_0$ and $c$ if they exist, hence it is guaranteed to terminate on boundedly periodic systems.

Finally, we highlight that the very same construction of \autoref{prop:ltl_cyclicity} can be used in a looping STS encoding instead of the plain one, allowing for the use of model-checking techniques that leverage lasso-shaped paths.

\subsection{Algorithms for TDLTL Model Checking}

The same encodings discussed in Definitions \ref{def:sts-basic} and \ref{def:sts-looping} can be used to solve the TDLTL model checking problem for MPL  systems using off-the-shelf model checking procedures. In fact, the two encodings capture the behavior of any MPL system as a STS (that is the natural input for existing model-checking tools such as \textsc{nuXmv} \cite{nuxmv}), and TDLTL can be captured in LTL by extending the STS with history variables tracking the evolution of the system.

More formally, given a TDLTL formula $\varphi$ we first identify all the terms $x_i^{(k)}$ appearing in the formula for some $i$ and some $k>0$. These are the variables that need to be monitored, because plain LTL does not allow for cross-time atoms. Monitoring a variable here means to create additional variables that ``guess'' the future values of $x_i$ by means of transition relation constraints. We first define a \emph{Looping STS Encoding with Monitors} as a Looping STS Encoding augmented with the monitor variables needed for a set of terms of the form $x_i^{(k)}$, then we show how to use such an encoding to reduce the TDLTL model-checking to plain LTL model-checking for infinite-state transition systems.

\begin{defn}[Looping STS Encoding with Monitors]\label{def:sts-looping-monitor}
    Given the MPL system \eqref{mpl} with an initial state constraint $\iota$ in \qflra and a set $M$ of terms of the form $x_i^{(k)}$ to be monitored, let  $\langle V, \iota, T \rangle$ be the Looping STS encoding for $A$ and $\iota$; the looping STS encoding with monitors is $\langle V', \iota, T' \rangle$, such that:
    \begin{itemize}
        \item $V \dot{=} V \cup \{m_{i,j} \mid x_i^{(k)} \in M, j\in[1,\cdots,k]\}$, where all the $m_{i,j}$ variables are of real type;
        \item $T'$ adds the following constraints to $T$: $\bigwedge_{x_i^{(k)} \in M} \left( (m_{i,1} = x'_i) \wedge \bigwedge_{j=2}^k (m_{i,j} = m'_{i,j-1}) \right)$.
    \end{itemize}
\end{defn}

Given a TDLTL formula $\varphi$, let $M_\varphi$ be the set of all the terms of the form $x_i^{(k)}$ appearing in the formula. Let $\tilde{\varphi}$ be the plain LTL formula where each such term is substituted with the corresponding variable $m_{i, k}$ in the Looping STS Encoding with Monitors, with monitor $M_\varphi$, namely: 
$$
\tilde{\varphi} \dot{=} \varphi[m_{i, k}/x_i^{(k)}]. 
$$
It is easy to see that $\tilde{\varphi}$ is syntactically an LTL formula (however, it is not a TDLTL one) as all the occurrences of TD atoms have been replaced by regular arithmetic differences over variables at the current time.

\begin{prop}\label{prop:tdltl_via smv}
    Given the MPL system \eqref{mpl} with an initial state constraint $\iota$ in \qflra and a TDLTL formula $\varphi$, $(A, \iota) \models \varphi$ if and only if $\tilde{\varphi}$ is  valid in the the Looping STS Encoding with Monitors for $A$, $\iota$ and $M_\varphi$.
\end{prop}
\begin{proof}
To show the proposition, it suffices to note that the semantics of TDLTL and LTL coincide,  with the exception that LTL does not allow for cross-time predicates. The rewriting in $\tilde{\varphi}$ is obviously correct if and only if, for any trace of the original MPL system, there exists a trace in the Looping STS Encoding with Monitors such that each monitor variable $m_{i, k}$ is equal to $x_i^{(k)}$ and vice-versa, for every trace of the Looping STS Encoding with Monitors each monitor variable $m_{i, k}$ is equal to $x_i^{(k)}$. This means that the original system and the encoding are in a bi-simulation relation.

This bi-simulation relation can be claimed by noting that the Looping STS Encoding with Monitors is in bi-simulation with the Looping STS Encoding, which in turn is a bi-simulation of the original MPL system. Moreover, we note that the variables $m_{i, k}$ faithfully predict the subsequent values of $x_i^{(k)}$ because the system is fully deterministic and entirely determined by the initial vector.
\end{proof}

\autoref{prop:tdltl_via smv} highlights a direct way of using any LTL model-checking procedure to verify TDLTL properties at the cost of adding additional variables to the STS, but without changing the modal structure of the TDLTL property. An alternative technique to use off-the-shelf model checkers for TDLTL, which will be formalized in the next section, is to exploit the determinism of MPL systems, to transform the TDLTL formula into an equi-satisfiable initial TDLTL formula, and to observe that any initial TDLTL formula coincides with a plain LTL formula that can be checked by standard tools. More formally, in the next section we will define a $\mathtt{get}\_\mathtt{initial}(A, \varphi)$ function that, for a specific MPL system with matrix $A$, transforms a possibly non-initial TDLTL formula $\varphi$ into an initial TDLTL formula. This transformation yields an equivalent TDLTL formula that is specific to a MPL system: thus, it can be fed to any appropriate LTL model checker, such as nuXmv.  

\section{Algorithms based on SMT}
\label{sec-algorithms-smt}
This section presents procedures to tackle  problems mentioned in Section \ref{sec-problem-definition} by applying SMT-solving. We first show that the syntax and operations of max-plus algebra can be expressed as a \qfrdl formula.
\begin{prop}[\cite{Mufid2020}]
\label{mpl_rdl_prop}
Given real-valued variables $\mathtt{x}_1,\ldots,\mathtt{x}_n$ and max-plus scalars $a_1,\ldots,a_n\in\Rmax$, the equation $\mathtt{x}^\prime = \max\{\mathtt{x}_1+a_1,\ldots,\mathtt{x}_n+a_n\}$ is equivalent to
\begin{equation}
    \bigwedge_{i=1}^n\left( \mathtt{x}^\prime\geq \mathtt{x}_i+a_i \right) \wedge \bigvee_{i=1}^n \left(\mathtt{x}^\prime= \mathtt{x}_i+a_i\right).
\end{equation}
\end{prop}
\noindent Consequently, by \autoref{mpl_rdl_prop}, any MPL system in \eqref{mpl} can be expressed as a \qfrdl formula as follows: 
\ifthenelse{\boolean{papertwocol}}{
\begin{align}
\bigwedge_{i=1}^n\Biggl(&\!\!\left(\bigwedge_{j\in \mathtt{fin}_i} \texttt{x}^{(k)}_i-\texttt{x}^{(k-1)}_j\geq A(i,j)\right)\wedge\nonumber\\
&\left(\bigvee_{j\in \mathtt{fin}_i} \texttt{x}^{(k)}_i-\texttt{x}^{(k-1)}_j=A(i,j)\right)\!\!\Biggr), 
\label{mpl_smt}
\end{align}

}{
\begin{equation}
\displaystyle\bigwedge_{i=1}^n\left(\!\!\left(\bigwedge_{j\in \mathtt{fin}_i} \texttt{x}^{(k)}_i-\texttt{x}^{(k-1)}_j\geq A(i,j)\right)\wedge\left(\bigvee_{j\in \mathtt{fin}_i} \texttt{x}^{(k)}_i-\texttt{x}^{(k-1)}_j=A(i,j)\right)\!\!\right), 
\label{mpl_smt}
\end{equation}
}
where 
$\mathtt{fin}_i$ is a set containing the indices of the finite elements of $A(i,\cdot)$. For the sake of simplicity, we denote \eqref{mpl_smt} as $\mathtt{SymbMPL}(A,\mathcal{V}^{(k-1)},\mathcal{V}^{(k)})$ where $\mathcal{V}^{(k)}=\{\mathtt{x}_1^{(k)},\ldots,\mathtt{x}_n^{(k)}\}$ is the set of (symbolic) variables encompassing the states of the MPL system in \eqref{mpl} at time horizon $k$. 

\begin{prop}[{\cite[Propositions 6-7]{Mufid2021}}]
\label{ori_to_reduced}
Given real-valued variables $\mathtt{x}_1,\ldots,\mathtt{x}_n$ and max-plus scalars $a_1,\ldots,a_n$, $b_1,\ldots,b_n\in \Rmax$, the inequality
\begin{equation}
F\equiv\bigoplus_{i=1}^p(\mathtt{x}_i+a_i)\sim \bigoplus_{j= 1}^p (\mathtt{x}_j+b_j)
\label{origin}
\end{equation} 
is equivalent to 
\begin{equation}
F^\ast\equiv\bigoplus_{i\in S_1}(\mathtt{x}_i+a_i)\sim \bigoplus_{j\in S_2} (\mathtt{x}_j+b_j),
\label{reduced}
\end{equation} 
where $S_1= \{1,\ldots,n\}\setminus\{1\leq k\leq n \mid  a_k= \varepsilon~\text{or}~ \neg (a_k\sim b_k)\}$ and $S_2=\{1,\ldots,n\} \setminus\{1\leq k\leq n \mid  b_k= \varepsilon~\text{or}~ a_k\sim b_k\}$, respectively. Furthermore, 
\ifthenelse{\boolean{papertwocol}}{
\begin{align}
F^\ast &\equiv \bigwedge_{j\in S_2 }\left( \bigvee_{i\in S_1 } (\texttt{x}_i-\texttt{x}_j\sim b_j-a_i)\right) \nonumber\\
&\equiv \bigvee_{i\in S_1 }\left( \bigwedge_{j\in S_2 } (\texttt{x}_i-\texttt{x}_j\sim b_j-a_i)\right).
\label{CNF_DNF}
\end{align}
}{
\begin{equation}
F^\ast\equiv \bigwedge_{j\in S_2 }\left( \bigvee_{i\in S_1 } (\texttt{x}_i-\texttt{x}_j\sim b_j-a_i)\right) \equiv
\bigvee_{i\in S_1 }\left( \bigwedge_{j\in S_2 } (\texttt{x}_i-\texttt{x}_j\sim b_j-a_i)\right).
\label{CNF_DNF}
\end{equation}
}
If $S_1=\emptyset$ then $F^\ast\equiv \texttt{false}$. On the other hand, if $S_2=\emptyset$ then $F^\ast\equiv \texttt{true}$. \QEDB
\end{prop}

Thanks to these results we can now define a transformation which, given an MPL system with a state matrix $A\in\Rmax^{n\times n}$ and a TD formula $f$, rewrites $f$ into an initial TD formula. We indicate such transformation  $\mathtt{get}\_\mathtt{initial}(A,f)$ and we define it in the following.
First, we translate all the TD propositions of $f$ in the form of $\texttt{x}^{(k)}_{i}-\texttt{x}^{(l)}_{j}\sim \alpha$ into the following inequality
\begin{equation}
\label{TD_prop2ineq}
\bigoplus_{r=1}^n \left(\texttt{x}_r +A^{\otimes k}(i,r)\right)\sim \bigoplus_{s=1}^n \left(\texttt{x}_s+\alpha+A^{\otimes l}(j,s)\right).  
\end{equation}
Then, we can translate $f$ into an initial TD formula by applying the rewriting \eqref{CNF_DNF} in \autoref{ori_to_reduced} for each inequality in \eqref{TD_prop2ineq}. We summarise the discussion in the following proposition and then present \autoref{ex:TD}. 



\begin{prop}[\cite{Mufid2021}]
Given a TD formula $f$ and $A\in \Rmax[n][n]$, let $g$ be $\mathtt{get}\_\mathtt{initial}(A,f)$. Then, for any orbit $\pi \in \mathsf{Orb}(A)$, $\pi \models f$ iff $\pi \models g$.
\label{non2formula}
\end{prop}

\begin{exmp}
\label{ex:TD}
Let us consider a TD formula $ f=\texttt{x}_1^{(1)}-\texttt{x}_1^{(0)}\leq 5$ for an MPL system in \autoref{ex1}. The formula represents ``the delay between the second and first departures of trains on station $S_1$ is no greater than 5 time unit''. It consists of one non-initial TD proposition. The translation of $f$ into an  initial TD formula is as follows: 
\ifthenelse{\boolean{papertwocol}}{%
\vskip -18pt
\begin{align*}
    \texttt{x}_1^{(0)}-\texttt{x}_1^{(1)}\geq -5 \Leftrightarrow&\texttt{x}_1+5\geq \max\{\texttt{x}_1+2,\texttt{x}_2+5\}\\
    \Leftrightarrow& \texttt{x}_1-\texttt{x}_2\geq 0,
\end{align*}%
}{
\[\begin{array}{lllll}
    \texttt{x}_1^{(0)}-\texttt{x}_1^{(1)}\geq -5&\Leftrightarrow&\texttt{x}_1+5\geq \max\{\texttt{x}_1+2,\texttt{x}_2+5\}&\Leftrightarrow& \texttt{x}_1-\texttt{x}_2\geq 0,\\
\end{array}\] 
}

Hence, we have $\mathtt{get}\_\mathtt{initial}(A,f)=\texttt{x}_1-\texttt{x}_2\geq 0$. Suppose we consider an orbit $\pi$ in \autoref{ex1}. It is easy to verify that $\pi\models \texttt{x}_1^{(1)}-\texttt{x}_1^{(0)}\leq 5$ since $x_1(1)-x_1(0)=5$. Similarly, $\pi\models \texttt{x}_1-\texttt{x}_2\geq 0$ since $x_1(0)-x_2(0)=0$. 
\QEDB
\end{exmp}

Note that this same rewriting allows the transformation of any TDLTL formula into an initial TD formula by simply transforming the TD-atoms appearing in the formula by means of $\mathtt{get}\_\mathtt{initial}$. Note that $\mathtt{get}\_\mathtt{initial}$ is an MPL-dependent transformation of a TD-formula, meaning that the MPL system dynamics are used to derive the rewritten formula. The resulting formula is therefore equivalent only on the specific MPL system (characterised by matrix $A$ in \eqref{mpl}) used for the transformation.

\subsection{Algorithms for Transient and Cyclicity Computation}
The computation of local transient and cyclicity w.r.t an initial vector $\textbf{x}(0)$ can be done easily by iterating \eqref{mpl}. Similarly, as per \autoref{transient_bound}, the global transient and cyclicity of $A\in\Rmax[n][n]$ can be obtained by computing the power of the matrix $A^{\otimes 0},A^{\otimes 1},\ldots$ until we find $l,c\geq 0$ such that
$
A^{\otimes(l+c)}=(\lambda\times c)\otimes A^{\otimes l}    
$, where $\lambda$ is the max-plus eigenvalue of $A$. As a result, for each $\textbf{x}\in \mathbb{R}^n$, we have
\begin{equation}
\label{eq_checking}
    A^{\otimes(l+c)}\otimes \textbf{x }=(\lambda\times c)\otimes A^{\otimes l} \otimes \textbf{x}.
\end{equation}
Instead of computing the power of the matrix, the SMT-based computation of transient and cyclicity is implemented by translating \eqref{eq_checking} into an SMT instance. First we denote $R=A^{\otimes (l+c)}$ and $S=(\lambda \times c)\otimes  A^{\otimes l}$. Then, \eqref{eq_checking} can be expressed as the conjunction of max-plus inequalities as follows: 
\ifthenelse{\boolean{papertwocol}}{
\begin{align}
\label{eq_func}
\begin{array}{l}
\displaystyle\bigwedge_{k=1}^n\left( \!\!\left(\bigoplus_{i=1}^n (\mathtt{x}_i+r_{ki})\geq \bigoplus_{j=1}^n (\mathtt{x}_j+s_{kj})\!\right)\right.\!\!\wedge  \\
\hspace*{4.5ex}\displaystyle \left.\left(\bigoplus_{i=1}^n (\mathtt{x}_i+s_{ki})\geq \bigoplus_{j=1}^n (\mathtt{x}_j+r_{kj})\right)\!\!\!\right)\!.
\end{array}
\end{align}
}{
\begin{equation}
\label{eq_func}
\bigwedge_{k=1}^n\!\left(\!\!\!\left(\bigoplus_{i=1}^n (\mathtt{x}_i+r_{ki})\geq \bigoplus_{j=1}^n (\mathtt{x}_j+s_{kj})\!\right)\!\!\wedge\!\!
\left(\bigoplus_{i=1}^n (\mathtt{x}_i+s_{ki})\geq \bigoplus_{j=1}^n (\mathtt{x}_j+r_{kj})\right)\!\!\!\right)\!.
\end{equation}
}
The scalar $r_{ki}$ (resp. $s_{ki}$) represents the element of $R$ (resp. $S$) at row $k$ and column $i$. For simplicity, we denote \eqref{eq_func} as $\mathtt{Eq}\mathtt{Func}(R,S)$. By \autoref{ori_to_reduced}, each conjunct in \eqref{eq_func} can be expressed as a \qfrdl formula. 

Algorithm \ref{comp_cyc_trans_smt} illustrates the SMT-based procedure to compute transient and cyclicity of an MPL system w.r.t a set of initial conditions $X\subseteq \mathbb{R}^n$. In this paper, we assume that the set $X$ is expressible as a \qflra formula. The algorithm starts by computing the cycle-time vector of the underlying MPL system. If the elements of cycle-time vector are not all equal, then we terminate the procedure since the transient does not exist (as per \autoref{periodic_everywhere}). Otherwise, we set the smallest possible values for the candidate of transient and cyclicity i.e., $l=0$ and $c=1$. Then, we generate the corresponding \qfrdl formula $F$ w.r.t. \eqref{eq_checking} in line 10. To check the validity of $F$, we use an SMT solver to check the unsatisfiability of its negation. If it is not satisfiable then the original formula is valid, and thus we obtain the transient and cyclicity from the current value of $l$ and $c$. Conversely, if it is satisfiable then there exists a counterexample, namely a falsifying formula $F$. We express the counterexample from a satisfying assignment of $ \neg F$ as a real-valued vector $v\in \mathbb{R}^n$ (line 15). This vector corresponds to the counterexample: its transient is greater than $l$ or its cyclicity does not divide $c$. The resulting local transient and cyclicity of $v$ are used to update the values for $(l,c)$ in lines 17-18. This process is repeated until either the SMT solver reports ``unsatisfiable'' in line 12, or $l+c$ exceeds the maximum bound $N$. We recall that the latter output may occur if the underlying MPL system is \textit{unboundedly periodic}.

\begin{algorithm}
\captionsetup{format=hang,width=\linewidth}
\caption{\footnotesize Computation of transient and cyclicity of $A$ w.r.t. a set of initial conditions $X$ via SMT-solving}
\label{comp_cyc_trans_smt}
\footnotesize
\begin{algorithmic}[1]
\Function{TransSMT}{$A,X,N$}
\State $\chi\gets \Call{CycleTimeVector}{A}$
\If{elements of $\chi$ are all equal}
\State $\lambda \gets \chi[0]$ \Comment{$\lambda$ is the eigenvalue of $A$}
\State{$n \gets $ \Call{Row}{$A$}} \Comment{number of rows of $A$}
 \For{$i \in \{1 \cdots n\}$}
       \State{$x[i] \gets $ \Call{MakeSMTRealVar}{\ }} \Comment{symbolic variables}
  \EndFor
\State $l\gets 0,c\gets 1$
\While{$(l+c)\leq N$}
\State $F \gets \mathtt{Eq}\mathtt{Func} (A^{\otimes l+c},(\lambda\times c)\otimes A^{\otimes l})$ 
\State $model \gets \Call{GetSMTModel}{X\wedge \neg F}$ 
\If{$model=\bot$} \Comment{formula is unsatisfiable}
\State \textbf{return} $\tuple{l,c}$
\Else \Comment{formula is satisfiable}
\State $v\gets\tuple{model(x[1]), \cdots, model(x[n])}$
\State{$\tuple{l^\prime, c^\prime} \gets $ \Call{TransCone}{$A$, $A^{\otimes l}\otimes v$, $N$}} 
\State $l\gets l+l^\prime$
\State $c\gets \texttt{LCM}(c,c^\prime)$
\EndIf
\EndWhile
\If{$((l+c)> N)$}
\State \textbf{print} ``terminated after reaching maximum bound''
\EndIf
\Else
\State \textbf{print} ``the transient does not exist''
\EndIf
\EndFunction
\end{algorithmic}
\end{algorithm}

\subsection{Algorithms for TDLTL Model Checking}
The procedure to verify \eqref{BMC_problem} exploits the fact the orbits of the MPL system \eqref{mpl} are potentially periodic and lasso-shaped. We build the algorithms based on the  Bounded Model Checking (BMC) technique \cite{Biere3,Biere1,Biere2}, which seeks to find the bounded counterexample of the given specification with a specific length $k$. As such, we generate an SMT instance corresponding to a bounded counterexample of the TDLTL formula $\varphi$. By \autoref{all_orbits}, such a counterexample can be generated over a lasso. First, we define the bounded version of \eqref{LTL_semantic} up to bound $k$ for a $(k,l)$-lasso $\pi$, as follows:
\ifthenelse{\boolean{papertwocol}}{
\begin{equation}
\left.
\label{LTL_semantic_lasso}
\begin{array}{ll}
\pi\models_k p& \text{iff}~ \pi[0]\models p,\\
\pi\models_k  \neg p& \text{iff}~ \pi\not\models_k p,\\
\pi\models_k \varphi_1\wedge \varphi_2& \text{iff}~\pi\models_k \varphi_1\wedge \pi\models_k\varphi_2,\\
\pi\models_k \varphi_1\vee \varphi_2& \text{iff}~\pi\models_k \varphi_1\vee \pi\models_k\varphi_2,\\
\pi\models_k \ltlX \varphi& \text{iff}~\pi[1..]\models_k \varphi,\\
\pi\models_k \varphi_1 \ltlU \varphi_2& \text{iff}~\exists 0\leq j\leq k.~\pi[j..]\models_k \varphi_2~\text{and}\\
&~~~\forall 0\leq i< j.~\pi[i..]\models_k \varphi_1,\\
\pi\models_k \varphi_1\ltlR\varphi_2& \text{iff}~\forall 0\leq j\leq k.~\pi[j]\models_k \varphi_2~\text{or}\\
&~~~\exists 0\leq i\leq k.~(\pi[i..]\models_k \varphi_1\wedge \\
&~~~\forall h\leq i.~ \pi[h..]\models_k \varphi_2 ),\\
 \pi\models_k  \ltlF \varphi& \text{iff}~\exists 0\leq j\leq k.~\pi[j..]\models_k \varphi,\\
 \pi\models_k  \ltlG \varphi& \text{iff}~\forall 0\leq j\leq k.~\pi[j..]\models_k \varphi.
\end{array}\!\right\}
\end{equation}
}
{
\begin{equation}
\left.
\label{LTL_semantic_lasso}
\begin{array}{ll}
\pi\models_k p& \text{iff}~ \pi[0]\models p,\\
\pi\models_k  \neg p& \text{iff}~ \pi\not\models_k p,\\
\pi\models_k \varphi_1\wedge \varphi_2& \text{iff}~\pi\models_k \varphi_1\wedge \pi\models_k\varphi_2,\\
\pi\models_k \varphi_1\vee \varphi_2& \text{iff}~\pi\models_k \varphi_1\vee \pi\models_k\varphi_2,\\
\pi\models_k \ltlX \varphi& \text{iff}~\pi[1..]\models_k \varphi,\\
\pi\models_k \varphi_1 \ltlU \varphi_2& \text{iff}~\exists 0\leq j\leq k.~\pi[j..]\models_k \varphi_2~\text{and}~\forall 0\leq i< j.~\pi[i..]\models_k \varphi_1,\\
\pi\models_k \varphi_1\ltlR\varphi_2& \text{iff}~\forall 0\leq j\leq k.~\pi[j]\models_k \varphi_2~\text{or}\\
&~~~~~\exists 0\leq i\leq k.~(\pi[i..]\models_k \varphi_1\wedge \forall h\leq i.~ \pi[h..]\models_k \varphi_2 ),\\
 \pi\models_k  \ltlF \varphi& \text{iff}~\exists 0\leq j\leq k.~\pi[j,,]\models_k \varphi,\\
 \pi\models_k  \ltlG \varphi& \text{iff}~\forall 0\leq j\leq k.~\pi[j..]\models_k \varphi.
\end{array}\!\right\}
\end{equation}
}

Notice that, for a $(k,l)$-lasso $\pi$, $\pi[(k+1)..]$ is similar to $\pi[l..]$. Thus, it is straightforward to see that $\pi\models_k \varphi$ implies $\pi\models \varphi$.

Following the bounded semantics in  \eqref{LTL_semantic_lasso}, we now will describe how to translate a bounded counterexample of a TDLTL formula into an SMT instance.  Suppose $\psi$ is the negation of $\varphi$ i.e. $\psi\equiv \neg \varphi$. We recall that $\varphi$ and $\psi$ are assumed to be in positive normal form \eqref{LTL_syntact}. The notation $_l[\psi]_k^m$ denotes the witness encoding of $\psi$ (equivalently, the counterexample encoding of $\varphi)$ at position $0\leq m\leq k$ over a $(k,l)$-lasso. Similar to the  description in \cite{Biere3}, the encoding can be formulated as follows:
\ifthenelse{\boolean{papertwocol}}{
\begin{align*}
_l[p]_k^m:=& p^{(m)}, \\
_l[\neg p]_k^m:=& \neg ( _l[p]_k^m), \\
_l[\psi_1 \wedge \psi_2]_k^m:=& _l[\psi_1]_k^m \wedge ~_l[\psi_2]_k^m,    \\
_l[\psi_1 \vee \psi_2]_k^m:=&~ _l[\psi_1]_k^m \vee ~_l[\psi_2]_k^m, \\
_l[\ltlX \psi]_k^m:=&\left\{\begin{array}{ll}
_l[\psi]_k^{m+1},&\text{if}~m<k\\
_l[\psi]_k^{l} ,&\text{otherwise}
\end{array}\right.\\
 _l[\psi_1 \ltlU \psi_2]_k^m:=&
   \displaystyle\bigvee_{j=m}^k\left( ~_l[\psi_2]_k^{j}\wedge \bigwedge_{n=m}^{j-1}~ _l[\psi_1]_k^{n} \right)\vee
   \\
   &\displaystyle\bigvee_{j=l}^{m-1}\left( {}_l[\psi_2]_k^{j}\wedge \bigwedge_{n=m}^{k} {}_l[\psi_1]_k^{n}\wedge \bigwedge_{n=l}^{j-1} {}_l[\psi_1]_k^{n} \right),&\hspace*{\hack}\\
   _l[\psi_1 \ltlR\psi_2]_k^m:=&
   \displaystyle\left(\bigwedge_{j=m^*}^k \hspace*{-1ex} {}_l[\psi_2]_k^j\right)\vee\bigvee_{j=m}^k\left(\!\! {}_l[\psi_1]_k^{j}\wedge\! \bigwedge_{n=m}^{j} \!\! {}_l[\psi_2]_k^{n} \right)\vee & \hspace*{\hack}\\
   &\displaystyle\bigvee_{j=l}^{m-1}\left( ~_l[\psi_1]_k^{j}\wedge \bigwedge_{n=m}^{k}~ _l[\psi_2]_k^{n}\wedge \bigwedge_{n=l}^{j}~ _l[\psi_2]_k^{n} \right),\\
   _l[\ltlG \psi]_k^m:=& \displaystyle \bigwedge_{j=m^*}^k ~ _l[\psi]_k^j,\\
   _l[\ltlF \psi]_k^m:=& \displaystyle \bigvee_{j=m^*}^k ~ _l[\psi]_k^j.
\end{align*}
where $m^*=\min\{m,l\}$ and $p$ is an initial TD proposition.
The final formula, which is satisfiable iff there exists a $(k,l)$-lasso $\pi$ such that $\pi\not\models_k \varphi$, is given by:
\begin{equation}
\label{witness}
\begin{array}{l}
\displaystyle
\bigwedge_{i=0}^k \mathtt{SymbMPL}(A,\mathcal{V}^{(i)},\mathcal{V}^{(i+1)})\wedge\\ \hspace*{10ex}\mathsf{Loop}(A,k+1,l,\lambda)\wedge {}_l[\psi]_k^0,
\end{array}
\end{equation}
}
{
\begin{align*}
_l[p]_k^m:=& p^{(m)},    &\hspace*{\hack}
_l[\neg p]_k^m:=& \neg ( _l[p]_k^m), \\
_l[\psi_1 \wedge \psi_2]_k^m:=& _l[\psi_1]_k^m \wedge ~_l[\psi_2]_k^m,    &\hspace*{\hack}
_l[\psi_1 \vee \psi_2]_k^m:=&~ _l[\psi_1]_k^m \vee ~_l[\psi_2]_k^m, \\
_l[\ltlX \psi]_k^m:=&\left\{\begin{array}{ll}
_l[\psi]_k^{m+1},&\text{if}~m<k\\
_l[\psi]_k^{l} ,&\text{otherwise}
\end{array}\right.& \hspace*{\hack}_l[\ltlF \psi]_k^m:=& \displaystyle \bigvee_{j=\min\{m,l\}}^k ~ _l[\psi]_k^j,\\
 _l[\psi_1 \ltlU \psi_2]_k^m:=&
   \displaystyle\bigvee_{j=m}^k\left( ~_l[\psi_2]_k^{j}\wedge \bigwedge_{n=m}^{j-1}~ _l[\psi_1]_k^{n} \right)\vee
   & \hspace*{\hack}_l[\ltlG \psi]_k^m:=& \displaystyle \bigwedge_{j=\min\{m,l\}}^k ~ _l[\psi]_k^j,
   \\
   &\displaystyle\bigvee_{j=l}^{m-1}\left( ~_l[\psi_2]_k^{j}\wedge \bigwedge_{n=m}^{k}~ _l[\psi_1]_k^{n}\wedge \bigwedge_{n=l}^{j-1}~ _l[\psi_1]_k^{n} \right),&\hspace*{\hack}\\
   _l[\psi_1 \ltlR\psi_2]_k^m:=&
   \displaystyle\left(\bigwedge_{j=\min\{m,l\}}^k {}_l[\psi_2]_k^j\right)\vee\bigvee_{j=m}^k\left( ~_l[\psi_1]_k^{j}\wedge \bigwedge_{n=m}^{j}~ _l[\psi_2]_k^{n} \right)\vee & \hspace*{\hack}\\
   &\displaystyle\bigvee_{j=l}^{m-1}\left( ~_l[\psi_1]_k^{j}\wedge \bigwedge_{n=m}^{k}~ _l[\psi_2]_k^{n}\wedge \bigwedge_{n=l}^{j}~ _l[\psi_2]_k^{n} \right),&\hspace*{\hack}
\end{align*}
where $p$ is an initial TD proposition.
The final formula, which is satisfiable iff there exists a $(k,l)$-lasso $\pi$ such that $\pi\not\models_k \varphi$, is given by:
\begin{eqnarray}
\label{witness}
\bigwedge_{i=0}^k \mathtt{SymbMPL}(A,\mathcal{V}^{(i)},\mathcal{V}^{(i+1)})\wedge \mathsf{Loop}(A,k+1,l,\lambda)\wedge~ _l[\psi]_k^0,
\end{eqnarray}
}
where $\lambda$ is the max-plus eigenvalue of $A$  
and $\mathsf{Loop}(A,k+1,l,\lambda)$ represents the looping constraint i.e., $\bigwedge_{i=1}^n (\mathtt{x}_i^{(k+1)}-\mathtt{x}_i^{(l)}=\lambda\times(k-l+1))$. Recall that, if the orbit of $\textbf{x}(0)$ w.r.t. $A$ is a $(k,l)$-lasso, then $\textbf{x}(k+1)=(\lambda\times (k-l+1))\otimes \textbf{x}(l)$. Furthermore, the first conjunct of \eqref{witness} corresponds to the executions of \eqref{mpl} up to bound $k$.

By the same procedure used in Proposition \ref{non2formula}, one can translate $\mathsf{Loop}(A,k+1,l,\lambda)\wedge{} _l[\psi]_k^0$ into an SMT formula over the variables $\mathcal{V}^{(0)}$ only (i.e., instead of representing the variables at each time in the orbit, we only define the formula over the initial state). 
Abusing the notation of TD formulae, we indicate the ``initialised'' version of \eqref{witness} as 
\begin{eqnarray}
\label{witness_initial}
\mathsf{get}\_\mathsf{initial}\left(A,\bigwedge_{i=0}^k  \mathsf{Loop}(A,k+1,l,\lambda)\wedge~ _l[\psi]_k^0\right).
\end{eqnarray}
The first conjunct of \eqref{witness} is not included because all TD propositions in \eqref{witness_initial} are initial ones. The number of TD propositions in \eqref{witness_initial} may be much larger compared to \eqref{witness}, especially for TDLTL formulae with multiple temporal operators. 
On the other hand, the encoding \eqref{witness_initial} has an advantage w.r.t. the number of variables: notice that \eqref{witness} consists of variables from $\mathcal{V}^{(0)}\cup\ldots\cup\mathcal{V}^{(k+1)}$, whereas  \eqref{witness_initial} involves $\mathcal{V}^{(0)}$ only. 

It has been shown in \cite{Mufid2021} that the completeness threshold to verify \eqref{BMC_problem} is determined by the pair of transient and cyclicity.
\begin{prop}[\cite{Mufid2021}]
\label{CT}
Given a periodic MPL system \eqref{mpl} with a set of initial conditions $X$ and a TDLTL formula $\varphi$, an upper bound of the completeness threshold to verify $\mathsf{Orb}(A,X)\models \varphi$ is given by $\mathsf{tr}(A,X)+\mathsf{cyc}(A,X)-1$. 
\end{prop}

In this paper, we present four variants of an algorithm, hinging on two independent factors.
The first factor is the computation of the bound, which could be carried out either \textit{incrementally} or \textit{upfront}. The other factor is the unrolling of the orbits of the MPL system, which can be either
\textit{unrolled} or \textit{initialised}. In the unrolled algorithms, we symbolically encode the orbit of MPL systems using \eqref{mpl_smt}. This encoding uses multiple sets of SMT variables up to the current bound. On the other hand, the initialised algorithms employ only one set of variables, thanks to some algebraic properties of MPL systems and \autoref{non2formula}. 

\subsection*{Incremental Approaches}
As mentioned before, orbits of MPL systems are potentially periodic with transient $l$ and cyclicity $c$. Hence, in incremental procedures, we search a finite counterexample of a TDLTL formula $\varphi$ with length $k$ in a shape of $(k,l)$-lasso, where initially we set $l=0$ and $c=1$ (the smallest possible values). From these values, we generate the looping constraint $\mathsf{Loop}(A,k+1,l,\lambda)$ and $_l[\neg \varphi]_k^0$, where $\lambda $ is the max-plus eigenvalue of $A$. The formula $X\wedge F$, with $F$ given by \eqref{witness} or \eqref{witness_initial}, corresponds to a counterexample of $\varphi$ i.e., a $(k,l)$-lasso $\pi\in \mathsf{Orb}(A,X)$,  such that $\pi\not\models\varphi$. 

We use an SMT solver to check the satisifiability of $X\wedge F$. If the SMT solver reports $\mathtt{SAT}$ then a counterexample is found. On the other hand, if the SMT solver reports $\mathtt{UNSAT}$, we increment the step bound. Instead of increasing the bound by one, we use the SMT solver to check whether there exists an orbit with larger transient $l^\prime$ or cyclicity $c^\prime$. The value $l^\prime+c^\prime-1$ then becomes the new bound. These two steps are repeated until either 1) a counterexample is found; 2) the bound cannot be incremented; or 3) the bound is deemed too large. The second outcome suggests that the specification is valid, since the bound exceeds the upper bound of the completeness threshold given by \autoref{CT}. For the last outcome, we use a large integer as a termination condition. 

The \textit{incremental} approaches are illustrated in \autoref{BMC}. We name the procedure that uses the encoding in \eqref{witness}  \textit{unrolled-incremental}, since the first conjunct of \eqref{witness} represents the execution of \eqref{mpl} up to bound $k$. On the other hand, the alternative procedure with the encoding in \eqref{witness_initial} is called \textit{initialised-incremental}, due to the translation from  \autoref{non2formula}.
\ifthenelse{\boolean{papertwocol}}{
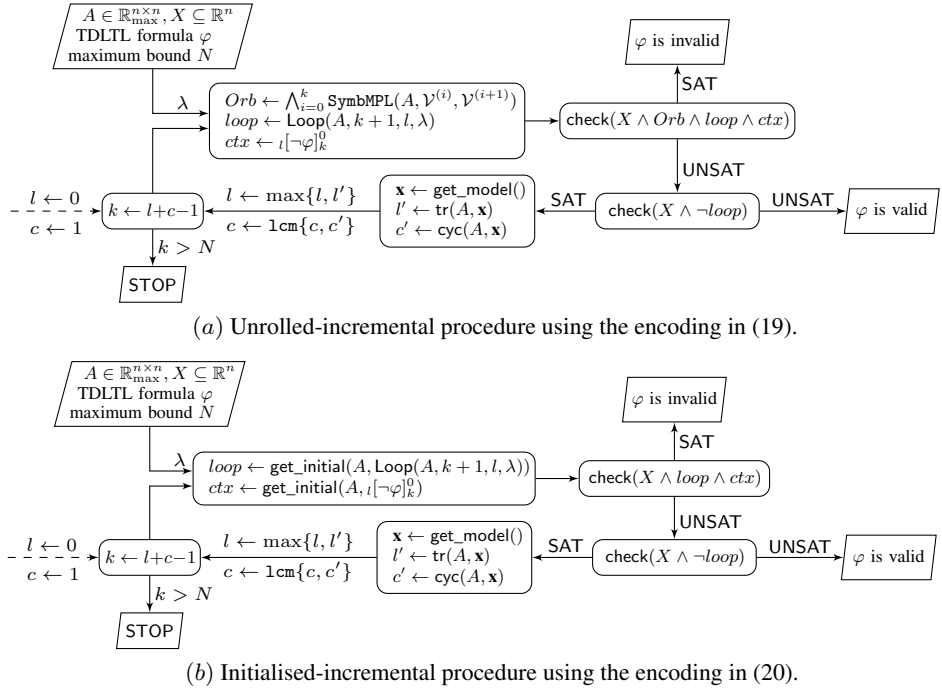
\begin{figure}[!ht]
\centering
\begin{footnotesize}
\begin{tikzpicture}[node distance = 1.5cm, auto,scale=1]
\node [par,text width=9em] (mpls) {
$\begin{array}{l}
A\in \mathbb{R}_{\max}^{n\times n},X\subseteq \mathbb{R}^n\\
\hspace*{-1.5ex}~\text{TDLTL formula}~ \varphi\\
\hspace*{-2.5ex}~\text{maximum bound}~N
\end{array}$
};    
\node [block,text width=17.5em,right of = mpls,xshift=1.5cm,yshift=-1.5cm] (mat) { 
\hspace*{-1ex}$\begin{array}{l}
Orb\gets \bigwedge_{i=0}^k \mathtt{SymbMPL}(A,\mathcal{V}^{(i)},\mathcal{V}^{(i+1)})\\
loop\gets \mathsf{Loop}(A,k+1,l,\lambda)\\
ctx \gets {}_l[\neg \varphi]_k^0
\end{array}$
};
\node [block,text width=12.5em,right of = mat,xshift=3.25cm] (SMT1) {$\mathsf{check}(X\wedge Orb \wedge loop\wedge ctx)$};
\node [par,text width=5em,above of = SMT1,yshift=-0.2cm] (out1) {$\varphi$ is invalid};    
\node [block,text width=8em,below of = SMT1,yshift=0.2cm] (SMT2) {\!$\mathsf{check}(X \wedge \neg  loop )$};
\node [par,text width=4.5em,below of = SMT2,yshift=0.2cm] (out2) {$\varphi$ is valid};    
\node [block,text width=7.4em,left of = SMT2,xshift=-1.6cm] (model) {
\hspace*{-1.5ex}$\begin{array}{l}
\textbf{x}\gets \mathsf{get}\_\mathsf{model}()\\
l^\prime \gets \mathsf{tr}(A,\textbf{x})\\
c^\prime \gets \mathsf{cyc}(A,\textbf{x})
\end{array}$
};    
\node [block,left of = model,text width=6em,xshift=-3.15cm] (tr) { $k\gets l+c-1$};    
\node [left of = tr] (emp) {};    
\node [par,below of = tr,text width=3em,minimum height=2em,yshift=0.2cm] (stop) { $\mathsf{STOP}$};
\path [line] (mpls) |- node [xshift=0.15cm,yshift=-0.05cm] {\scriptsize $\lambda$} ($(mat.west)+(0,0.1)$); 
\path [line,dashed] (emp) -- node [xshift=0cm,yshift=-0.5cm] {\scriptsize 
$\begin{array}{l}
l\gets 0\vspace*{1ex}\\
c\gets 1
\end{array}$
}(tr); 
\path [line] (mat) -- (SMT1); 
\path [line] (SMT1) -- node [xshift=0.65cm,yshift=0cm] {\scriptsize $\mathsf{SAT}$}(out1); 
\path [line] (SMT1) -- node [xshift=-0.05cm,yshift=-0.05cm] {\scriptsize $ \mathsf{UNSAT}$}(SMT2); 
\path [line] (SMT2) -- node [xshift=0cm,yshift=-0.05cm] {\scriptsize $\mathsf{UNSAT}$}(out2); 
\path [line] (SMT2) -- node [xshift=0.05cm,yshift=0.35cm] {\scriptsize $\mathsf{SAT}$}(model); 
\path [line] (model) -- node [xshift=0.cm,yshift=0.45cm] {\scriptsize 
$\begin{array}{l}
l\gets \max\{l,l^\prime\}\vspace*{1ex}\\
c\gets \mathtt{lcm}\{c,c^\prime\}
\end{array}$
}(tr); 
\path [line] (tr) |- ($(mat.west)+(0,-0.1)$); 
\path [line] (tr) -- node[xshift=-0.05cm] {\scriptsize $k> N$} (stop);
\draw node[below = of model,xshift=-0.6cm,yshift=0.5cm]{$(a)$ Unrolled-incremental procedure using the encoding in \eqref{witness}.};
\end{tikzpicture}\vspace*{1ex}\\
\begin{tikzpicture}[node distance = 1.5cm, auto,scale=1]
\node [par,text width=8.5em] (mpls) {
$\begin{array}{l}
A\in \mathbb{R}_{\max}^{n\times n},X\subseteq \mathbb{R}^n\\
\hspace*{-1.5ex}~\text{TDLTL formula}~ \varphi\\
\hspace*{-2.5ex}~\text{maximum bound}~N
\end{array}$
};    
\node [block,text width=18.5em,right of = mpls,xshift=1.8cm,yshift=-1.25cm] (mat) { \hspace*{-2ex}
$\begin{array}{l}
loop\gets \mathsf{get}\_\mathsf{initial}(A,\mathsf{Loop}(A,k+1,l,\lambda))\\
ctx \gets \mathsf{get}\_\mathsf{initial}(A,{}_l[\neg \varphi]_k^0)
\end{array}$
};
\node [block,text width=10em,right of = mat,xshift=3.1cm] (SMT1) {$\mathsf{check}(X\wedge loop\wedge ctx)$};
\node [par,text width=5em,above of = SMT1,yshift=-0.2cm] (out1) {$\varphi$ is invalid};    
\node [block,text width=8.5em,below of = SMT1,yshift=0.2cm] (SMT2) {$\mathsf{check}(X \wedge \neg  loop )$};
\node [par,text width=4.5em,below of = SMT2,yshift=0.2cm] (out2) {$\varphi$ is valid};    
\node [block,text width=7.5em,left of = SMT2,xshift=-1.8cm] (model) {\hspace*{-2ex}
$\begin{array}{l}
\textbf{x}\gets \mathsf{get}\_\mathsf{model}()\\
l^\prime \gets \mathsf{tr}(A,\textbf{x})\\
c^\prime \gets \mathsf{cyc}(A,\textbf{x})
\end{array}$
};    
\node [block,left of = model,text width=6em,xshift=-3.1cm] (tr) { $k\gets l+c-1$};    
\node [left of = tr] (emp) {};    
\node [par,below of = tr,text width=3em,minimum height=2em,yshift=0.2cm] (stop) { $\mathsf{STOP}$};
\path [line] (mpls) |- node [xshift=0.25cm,yshift=-0.05cm] {\scriptsize $\lambda$} ($(mat.west)+(0,0.1)$); 
\path [line,dashed] (emp) -- node [xshift=0cm,yshift=-0.5cm] {\scriptsize 
$\begin{array}{l}
l\gets 0\vspace*{1ex}\\
c\gets 1
\end{array}$
}(tr); 
\path [line] (mat) -- (SMT1); 
\path [line] (SMT1) -- node [xshift=0.65cm,yshift=0cm] {\scriptsize $\mathsf{SAT}$}(out1); 
\path [line] (SMT1) -- node [xshift=-0.05cm,yshift=-0.05cm] {\scriptsize $ \mathsf{UNSAT}$}(SMT2); 
\path [line] (SMT2) -- node [xshift=0cm,yshift=-0.05cm] {\scriptsize $\mathsf{UNSAT}$}(out2); 
\path [line] (SMT2) -- node [xshift=0.05cm,yshift=0.35cm] {\scriptsize $\mathsf{SAT}$}(model); 
\path [line] (model) -- node [xshift=0.cm,yshift=0.45cm] {\scriptsize 
$\begin{array}{l}
l\gets \max\{l,l^\prime\}\vspace*{1ex}\\
c\gets \mathtt{lcm}\{c,c^\prime\}
\end{array}$
}(tr); 
\path [line] (tr) |- ($(mat.west)+(0,-0.1)$); 
\path [line] (tr) -- node[xshift=-0.05cm] {\scriptsize $k> N$} (stop);
\draw node[below = of model,xshift=-0.45cm,yshift=0.6cm]{$(b)$ Initialised-incremental procedure using the encoding in \eqref{witness_initial}.};
\end{tikzpicture}
\end{footnotesize}
\captionsetup{format=hang,width=\linewidth}
\caption{Incremental Approaches. The function $\mathsf{check}$ is implemented in an SMT solver. The integer $N$ represents the allowed maximum bound.}
\label{BMC}
\end{figure}
}
{
\begin{figure}[!ht]
\centering
\begin{footnotesize}
\begin{tikzpicture}[node distance = 2cm, auto,scale=1]
\node [par,text width=9em] (mpls) {
$\begin{array}{l}
A\in \mathbb{R}_{\max}^{n\times n},X\subseteq \mathbb{R}^n\\
\hspace*{-1.5ex}~\text{TDLTL formula}~ \varphi\\
\hspace*{-2.5ex}~\text{maximum bound}~N
\end{array}$
};    
\node [block,text width=17em,right of = mpls,xshift=1.8cm,yshift=-1.5cm] (mat) { 
$\begin{array}{l}
Orb\gets \bigwedge_{i=0}^k \mathtt{SymbMPL}(A,\mathcal{V}^{(i)},\mathcal{V}^{(i+1)})\\
loop\gets \mathsf{Loop}(A,k+1,l,\lambda)\\
ctx \gets {}_l[\neg \varphi]_k^0
\end{array}$
};
\node [block,text width=12.5em,right of = mat,xshift=3.5cm] (SMT1) {$\mathsf{check}(X\wedge Orb \wedge loop\wedge ctx)$};
\node [par,text width=5em,above of = SMT1,yshift=-0.6cm] (out1) {$\varphi$ is invalid};    
\node [block,text width=8.5em,below of = SMT1,yshift=0.4cm] (SMT2) {$\mathsf{check}(X \wedge \neg  loop )$};
\node [par,text width=4.5em,right of = SMT2,xshift=1.8cm] (out2) {$\varphi$ is valid};    
\node [block,text width=8em,left of = SMT2,xshift=-1.9cm] (model) {
$\begin{array}{l}
\textbf{x}\gets \mathsf{get}\_\mathsf{model}()\\
l^\prime \gets \mathsf{tr}(A,\textbf{x})\\
c^\prime \gets \mathsf{cyc}(A,\textbf{x})
\end{array}$
};    
\node [block,left of = model,text width=5em,xshift=-3.4cm] (tr) { $k\gets l+c-1$};    
\node [left of = tr] (emp) {};    
\node [par,below of = tr,text width=3em,minimum height=2em,yshift=0.7cm] (stop) { $\mathsf{STOP}$};
\path [line] (mpls) |- node [xshift=0.4cm,yshift=-0.05cm] {\scriptsize $\lambda$} ($(mat.west)+(0,0.1)$); 
\path [line,dashed] (emp) -- node [xshift=0cm,yshift=-0.5cm] {\scriptsize 
$\begin{array}{l}
l\gets 0\vspace*{1ex}\\
c\gets 1
\end{array}$
}(tr); 
\path [line] (mat) -- (SMT1); 
\path [line] (SMT1) -- node [xshift=0.65cm,yshift=0cm] {\scriptsize $\mathsf{SAT}$}(out1); 
\path [line] (SMT1) -- node [xshift=-0.05cm,yshift=-0.05cm] {\scriptsize $ \mathsf{UNSAT}$}(SMT2); 
\path [line] (SMT2) -- node [xshift=0cm,yshift=-0.05cm] {\scriptsize $\mathsf{UNSAT}$}(out2); 
\path [line] (SMT2) -- node [xshift=0.05cm,yshift=0.35cm] {\scriptsize $\mathsf{SAT}$}(model); 
\path [line] (model) -- node [xshift=0.cm,yshift=0.45cm] {\scriptsize 
$\begin{array}{l}
l\gets \max\{l,l^\prime\}\vspace*{1ex}\\
c\gets \mathtt{lcm}\{c,c^\prime\}
\end{array}$
}(tr); 
\path [line] (tr) |- ($(mat.west)+(0,-0.1)$); 
\path [line] (tr) -- node[xshift=-0.05cm] {\scriptsize $k> N$} (stop);
\draw node[below = of model,xshift=0.5cm,yshift=1.2cm]{$(a)$ Unrolled-incremental procedure using the encoding in \eqref{witness}.};
\end{tikzpicture}\vspace*{1ex}\\
\begin{tikzpicture}[node distance = 2cm, auto,scale=1]
\node [par,text width=8.5em] (mpls) {
$\begin{array}{l}
A\in \mathbb{R}_{\max}^{n\times n},X\subseteq \mathbb{R}^n\\
\hspace*{-1.5ex}~\text{TDLTL formula}~ \varphi\\
\hspace*{-2.5ex}~\text{maximum bound}~N
\end{array}$
};    
\node [block,text width=18.5em,right of = mpls,xshift=1.8cm,yshift=-1.5cm] (mat) { 
$\begin{array}{l}
loop\gets \mathsf{get}\_\mathsf{initial}(A,\mathsf{Loop}(A,k+1,l,\lambda))\\
ctx \gets \mathsf{get}\_\mathsf{initial}(A,{}_l[\neg \varphi]_k^0)
\end{array}$
};
\node [block,text width=10em,right of = mat,xshift=3.5cm] (SMT1) {$\mathsf{check}(X\wedge loop\wedge ctx)$};
\node [par,text width=5em,above of = SMT1,yshift=-0.6cm] (out1) {$\varphi$ is invalid};    
\node [block,text width=8.5em,below of = SMT1,yshift=0.6cm] (SMT2) {$\mathsf{check}(X \wedge \neg  loop )$};
\node [par,text width=4.5em,right of = SMT2,xshift=1.8cm] (out2) {$\varphi$ is valid};    
\node [block,text width=8em,left of = SMT2,xshift=-1.9cm] (model) {
$\begin{array}{l}
\textbf{x}\gets \mathsf{get}\_\mathsf{model}()\\
l^\prime \gets \mathsf{tr}(A,\textbf{x})\\
c^\prime \gets \mathsf{cyc}(A,\textbf{x})
\end{array}$
};    
\node [block,left of = model,text width=5em,xshift=-3.4cm] (tr) { $k\gets l+c-1$};    
\node [left of = tr] (emp) {};    
\node [par,below of = tr,text width=3em,minimum height=2em,yshift=0.7cm] (stop) { $\mathsf{STOP}$};
\path [line] (mpls) |- node [xshift=0.4cm,yshift=-0.05cm] {\scriptsize $\lambda$} ($(mat.west)+(0,0.1)$); 
\path [line,dashed] (emp) -- node [xshift=0cm,yshift=-0.5cm] {\scriptsize 
$\begin{array}{l}
l\gets 0\vspace*{1ex}\\
c\gets 1
\end{array}$
}(tr); 
\path [line] (mat) -- (SMT1); 
\path [line] (SMT1) -- node [xshift=0.65cm,yshift=0cm] {\scriptsize $\mathsf{SAT}$}(out1); 
\path [line] (SMT1) -- node [xshift=-0.05cm,yshift=-0.05cm] {\scriptsize $ \mathsf{UNSAT}$}(SMT2); 
\path [line] (SMT2) -- node [xshift=0cm,yshift=-0.05cm] {\scriptsize $\mathsf{UNSAT}$}(out2); 
\path [line] (SMT2) -- node [xshift=0.05cm,yshift=0.35cm] {\scriptsize $\mathsf{SAT}$}(model); 
\path [line] (model) -- node [xshift=0.cm,yshift=0.45cm] {\scriptsize 
$\begin{array}{l}
l\gets \max\{l,l^\prime\}\vspace*{1ex}\\
c\gets \mathtt{lcm}\{c,c^\prime\}
\end{array}$
}(tr); 
\path [line] (tr) |- ($(mat.west)+(0,-0.1)$); 
\path [line] (tr) -- node[xshift=-0.05cm] {\scriptsize $k> N$} (stop);
\draw node[below = of model,xshift=0.5cm,yshift=1.2cm]{$(b)$ Initialised-incremental procedure using the encoding in \eqref{witness_initial}.};
\end{tikzpicture}
\end{footnotesize}
\captionsetup{format=hang,width=\linewidth}
\caption{Incremental Approaches. The function $\mathsf{check}$ is implemented in an SMT solver. The integer $N$ represents the allowed maximum bound.}
\label{BMC}
\end{figure}
}

\subsection*{Upfront Approaches}
Upfront approaches exploit the fact that the upper bound of the completeness threshold in \autoref{CT} is unrelated to the TDLTL formula $\varphi$, and that it can be computed via the SMT-based approach in Algorithm \ref{comp_cyc_trans_smt}. Hence, the upfront versions of the  procedures in \autoref{BMC} is obtained by generating \eqref{witness} or \eqref{witness_initial} with $l=
\mathsf{tr}(A,X)$ and $k=\mathsf{tr}(A,X)+\mathsf{cyc}(A,X)-1$. Together with the set of initial conditions $X$, we check the satisfaction of the resulting SMT instance. If it is satisfiable, then $\varphi$ is invalid. On the other hand, if it is unsatisfiable, then $\varphi$ is valid. The resulting procedures are then called \textit{unrolled-upfront} and \textit{initalised-upfront}, respectively.

\section{Related Work}
\label{sec-relatedwork}



Equipped with the notions introduced in the previous sections, we now relate our contributions to cognate works. 

The classical analysis of MPL systems depends on the algebraic properties of the state matrices such as eigenvalues and eigenvectors \cite{Subiono}, irreducibility \cite{Baccelli}, cycle-time vector \cite{Heidergott} and periodic behaviors \cite{Bart2}. Those properties can be studied by using the spectral theory of the matrices. The periodic behavior of MPL systems is particularly important as it defines the steady-state condition (in max-plus algebraic sense), starting from the bound called the transient, and with a period called cyclicity. The classical method to compute the transient and the cyclicity is by simply computing the power of the underlying state matrix as described in  \autoref{transient_bound}. The main drawback of this approach is that it can only be applied for an MPL system where the set of initial conditions is a max-plus cone. The graph-based approaches for transient computation are presented in \cite{Charron,Merlet,Nowak,Gerardo}. Yet, they only provide the upper bounds which are much larger than the actual transient.

Regarding the formal analysis for MPL systems, it has been first discussed in \cite{Dieky1} by employing the \emph{abstraction} procedure \cite{Baier,Graf}. Generally speaking, abstractions are techniques to generate a finite and smaller model from a large or even infinite (e.g., continuous state-space) model. In the case of MPL systems, the dynamics are transformed into an equivalent Piecewise Affine (PWA) model \cite{Heemels}. This allows to partition the state space of MPL systems into abstract states using Difference Bound Matrices (DBM) \cite{Mufid2018,Dill} as the underlying data structure; the MPL dynamics are then reduced to transitions among abstract states. This abstraction-based technique allows to verify specifications over the abstract model: if the specification holds on the abstract model, then it is also valid for the original MPL system \cite{Dieky1, Baier}. However, the invalidity of specification on the abstract model does not necessarily imply the same conclusion on the MPL system.

An abstraction-refinement procedure is then proposed in \cite{Dieky1}: the abstract model can be refined, so that it has a bisimulation relation \cite[Definition 7.1]{Baier} with the original MPL system. This means that the specification is true on MPL system if and only if it holds on the bisimilar abstract model. Unfortunately, the refinement procedure in general does not terminate, even for irreducible MPL systems. Furthermore, the surveyed techniques, whilst formal, suffer from the curse of dimensionality, since the abstraction computation runs on O$(n^{(n+3)})$ complexity where $n$ is the dimension of the state matrix.

The recent work \cite{Mufid2019} applies an alternative approach to verify MPL systems. A set of predicates is used to generate an abstraction of an MPL system. Predicates are automatically generated from the state matrix, as well as from the specifications under consideration. This \emph{predicate abstraction} of MPL systems, reminiscent of \cite{Graf,Clarke2}, is shown to be more scalable than the PWA-based abstraction in \cite{Dieky1}. A standard BMC procedure is then applied to verify the given specifications.   
It also has been proved in \cite[Lemma 2]{Mufid2019} that the completeness threshold for this BMC procedure is determined by the pair of transient and cyclicity. Furthermore, unlike in \cite{Dieky1}, this procedure does not require to generate the bisimilar abstract model. Instead, one can first check whether the counterexample (the behavior that invalidates the specification) is spurious: if it is not spurious, then the specification is indeed not valid on the MPL system; if instead a spurious counterexample is found, the abstraction of the MPL system can be refined using the procedure in \cite{Dieky3}, combined with standard lazy abstraction \cite{Henzinger}.

Despite successive ameliorations, the main drawback of the aforementioned formal methods is their scalability, as they can only be applied to MPL systems with relatively few variables (the dimension $n$ of vector $\textbf{x}$ in this work). There are a few elements contributing to the computational bottleneck (time and memory requirements) of these approaches. First, the worst-case complexity to generate the abstraction of $n$-dimensional MPL systems is O$(n^{n+3})$ \cite{Dieky1}. As a result, the number of abstract states grows exponentially, as $n$ increases. Second, the refinement procedures in \cite{Dieky1,Mufid2019} potentially lead to state-explosion problems. Another disadvantage is the limitations related to utilising the DBM data structure: in \cite{Mufid2019}, each proposition in the form of $\texttt{x}_i-\texttt{x}_j\sim c$ where $1\leq i,j\leq n,\sim\hspace*{0.5ex}\in\{>,\geq\}$ and $\alpha\in \mathbb{R}$ is transformed into a DBM in $\mathbb{R}^n$. As such, the more propositions are in an LTL formula, the more DBMs are needed, and therefore the larger is number of generated abstract states.

\section{Experimental Evaluation}
\label{sec-experimental-evaluation}

\subsection{Transient and Cyclicity}
We compare the performance of matrix-multiplication  (\autoref{comp_cyc_trans}), STS model checking (\autoref{fig:MC-TCC}), and SMT-based (\autoref{comp_cyc_trans_smt}) techniques, all employed to compute the transient of MPL systems. The experiments for matrix multiplication and SMT procedures are implemented in Python, while we perform an encoding into the SMV language and use the \textsc{nuXmv} tool \cite{nuxmv} or the STS model checking approach. For the SMT solver, we use Yices 2.2 \cite{Yices}. The computational benchmark has been implemented on an Intel\textregistered{} Xeon\textregistered{} CPU E5-1660 v3, 16 cores, 3.0GHz each, and 16GB of RAM. For the experiments, we generate 1000 irreducible matrices of dimension $n$, with $m$ finite elements in each row, where the values of the finite elements are rational numbers $\frac{p}{q}$ with $1\leq p\leq 100$ and $1\leq q\leq 5$. The locations of the finite elements are chosen randomly. We focus on irreducible matrices in order to ensure the termination of the algorithms. The matrix-multiplication technique (\autoref{comp_cyc_trans}) is initialised by setting $U$ to be a max-plus identity matrix, while for \autoref{comp_cyc_trans_smt} the set of initial conditions is expressed as $X\equiv \top$. For all experiments, we choose $N=10000$ as the maximum bound.
\ifthenelse{\boolean{papertwocol}}{
\begin{figure*}[!ht]
  \begin{subfigure}[t]{0.33\textwidth}
  \centering
  \includegraphics[width=.95\textwidth]{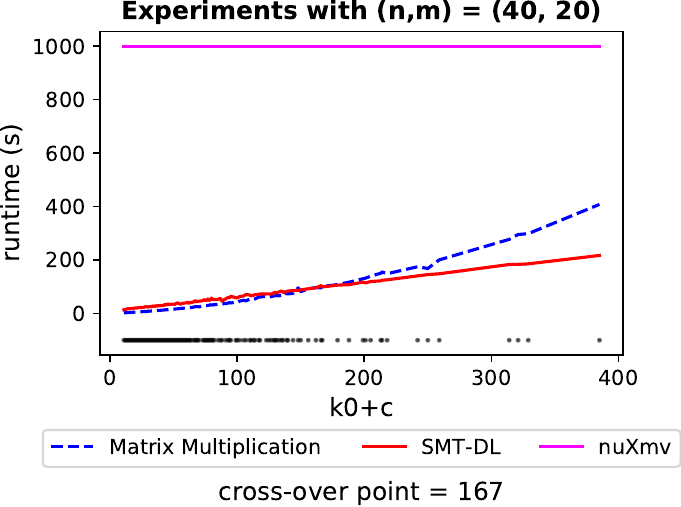}
  \caption{}
\end{subfigure}
\hspace{-2.ex}
\begin{subfigure}[t]{0.33\textwidth}
  \centering
  \includegraphics[width=.95\textwidth]{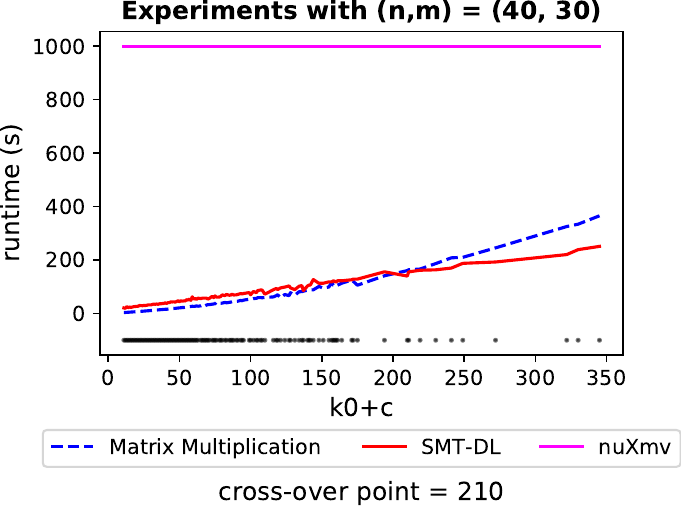}
  \caption{}
\end{subfigure}
\hspace{-2.ex}
\begin{subfigure}[t]{0.33\textwidth}
  \centering
  \includegraphics[width=.95\textwidth]{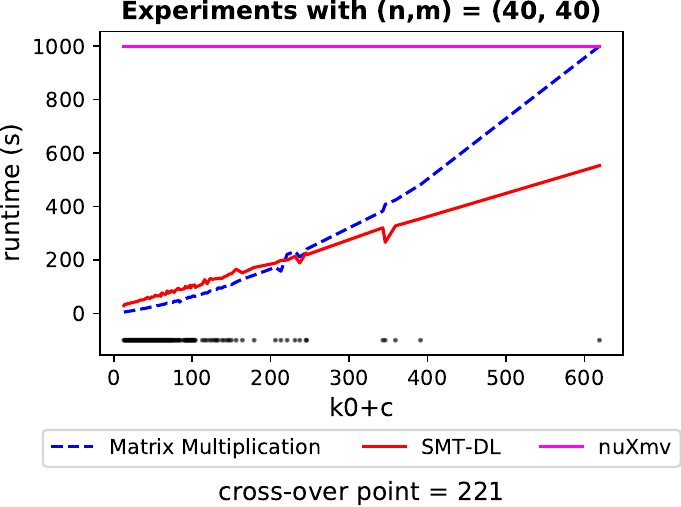}
  \caption{}
\end{subfigure}
    \captionsetup{format=hang,width=0.9\linewidth}
    \caption{The plots of running time of the \autoref{comp_cyc_trans} and  \autoref{comp_cyc_trans_smt} from 1000 experiments with $n=40$ and $m\in \{20,30,40\}$. A ``cross-over point'' is the smallest value of $k_0+c$ when  Algorithm \ref{comp_cyc_trans_smt} is faster.}
    \label{benchmark}
\end{figure*}
}
{
\begin{figure}[!ht]
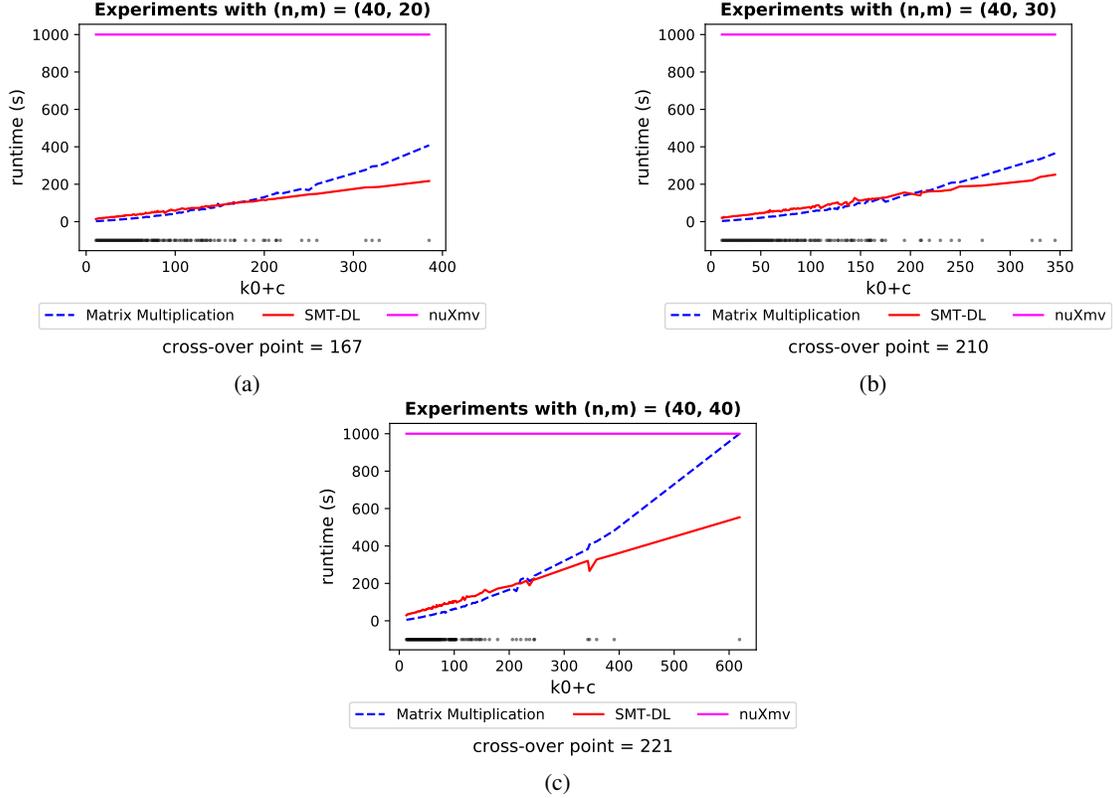

  \begin{subfigure}[t]{0.5\textwidth}
  \centering
  \includegraphics[scale=0.55]{plots/k0plusc_40_20.pdf}
  \caption{}
\end{subfigure}
\begin{subfigure}[t]{0.5\textwidth}
  \centering
  \includegraphics[scale=0.55]{plots/k0plusc_40_30.pdf}
  \caption{}
\end{subfigure}
\vspace*{2ex}
\begin{subfigure}[t]{1\textwidth}
  \centering
  \includegraphics[scale=0.55]{plots/k0plusc_40_40.pdf}
  \caption{}
\end{subfigure}
    \captionsetup{format=hang,width=\linewidth}
    \caption{The plots of running time of the matrix-multiplication technique and  \ref{comp_cyc_trans_smt} from 1000 experiments with $n=40$ and $m\in \{20,30,40\}$. A ``cross-over point'' is the smallest value of $k_0+c$ when  Algorithm \ref{comp_cyc_trans_smt} is faster.}
    \label{benchmark}
\end{figure}
}

\autoref{benchmark}(a)-(c) illustrate the experiments for $n=40$ and $m\in\{20,30,40\}$ (we report all the experiments in the extended version of this paper). They show the plots of the running times of Algorithm \ref{comp_cyc_trans} (dashed lines) and of Algorithm \ref{comp_cyc_trans_smt} (solid lines) against the resulting transient $k_0$ and cyclicity $c$ - the scattered plots (in black) correspond to the resulting $k_0+c$.
If there are several experiments with the same value of $k_0+c$, then we display the average running time among those experiments. The STS model checking approach timed out at 1000 seconds in all these experiments (it correctly terminates for smaller values of $n$).

With regards to the running time, the matrix-multiplication algorithm is faster when the values of $k_0+c$ are quite small. On the other hand, the larger is the value of $k_0+c$, the better is the performance of the SMT-based algorithm. We argue that this is because in Algorithm \ref{comp_cyc_trans_smt} there may be a large increment from the current guess of transient and cyclicity to the new ones. Whereas in Algorithm \ref{comp_cyc_trans}, the next candidate of transient and cyclicity is increased by one at each iteration.


{It is also evident that, from \autoref{benchmark}(a)-(c), the algorithm based on the STS model checking is the least efficient one. In fact, all experiments do not terminate after 1000 seconds. The main reason of this defect is that the STS encoding of MPL systems for transient and cyclicity checking(\autoref{def:sts-encoding-trans-cyc} reduces to a series of infinite-state model-checking problems (which are in general undecidable), while the SMT-based encoding reduces to a series of much cheaper, NP-hard checks.
\ifthenelse{\boolean{papertwocol}}{\textcolor{red}{\bf TODO: cite arXiv version here}}{
In Appendix \ref{sec-more experiments}, we present the experiments with smaller dimensions. Yet, the STS-based technique remains the slowest one.}}

\subsection{TDLTL Model Checking}
For the TDLTL Model Checking experiments, we generate 20 irreducible matrices of dimension $n\in\{4,6,8,10\}\cup \{12,16,\ldots,40\}$ with $ \frac{n}{2}$ finite elements in each row. The values of the finite elements are integers between 1 and 20, where the locations of the finite elements at each row are chosen randomly. We focus on irreducible matrices to ensure the termination of the procedures. With regards to the specifications, for each $n$ we generate randomly 20 TDLTL formulae where the propositions are in the form of $\texttt{x}_i-\texttt{x}_j\sim \alpha$, $i,j\in\{1,\ldots,n\},\sim \hspace*{0.5ex}\in \{>,\geq\}$, and $\alpha$ is an integer within the interval $[-20,20]$.
The randomised TDLTL formulae are generated using Spot \cite{Duret}.

The set of initial conditions for each experiment is $X=\mathbb{R}^n$. The experiments are implemented for each pair of matrix and specification. Thus, there are $20\times 20\times 2$ experiments for each $n$. We set 30 minutes as a $\mathtt{timeout}$ limit. {
For the experiments, we use three types of algorithms: (i) STS-based (Section \ref{sec-algorithms-model-checking}) , (ii) SMT-based (Section
\ref{sec-algorithms-smt}) and (iii) abstraction-based \cite{Dieky1}.}

\autoref{BMC_plot1} illustrates the performance comparison of abstraction-based, unrolled-incremental, and STS model checking (indicated as IC3-based in the plots as we use the IC3-IA\cite{ic3ia} algorithm of \textsc{nuXmv} to solve the model checking problem) algorithms for $n\in \{4,6,8,10\}$. These three algorithms are similar in the sense that they unroll the dynamics of MPL systems up to the underlying step bound. The scattered plots (in logarithmic scale) represent the running times (in second) for a pair of algorithms. It is clear that the abstraction-based algorithm is in general the least efficient one.
As expected, the dimension of MPL systems heavily affects the runtime of the abstraction-based procedure: 
all experiments for 10-dimensional MPL systems yield $\mathtt{timeout}$. For this reason, we do not pursue the abstraction-based experiments for higher dimensions.

\ifthenelse{\boolean{papertwocol}}
{
\begin{figure*}[!ht]
  \centering
  \begin{subfigure}[!ht]{\textwidth}
    \centering
    \includegraphics[scale=0.45]{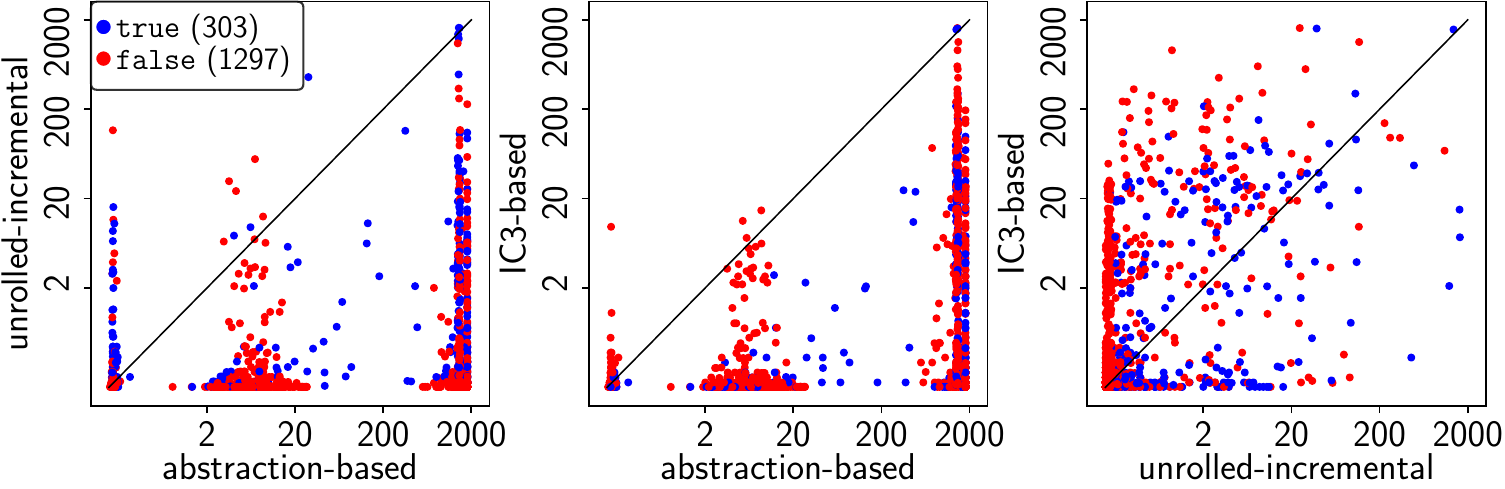}
  \end{subfigure}
  \captionsetup{format=hang,width=\linewidth}
    \caption{Comparison of abstraction-based, unrolled-incremental and IC3-based algorithms. 
    }
    \label{BMC_plot1}
\end{figure*}
}
{
\begin{figure}[!ht]
  \centering
  \begin{subfigure}[!ht]{\textwidth}
    \centering
    \includegraphics[scale=0.55]{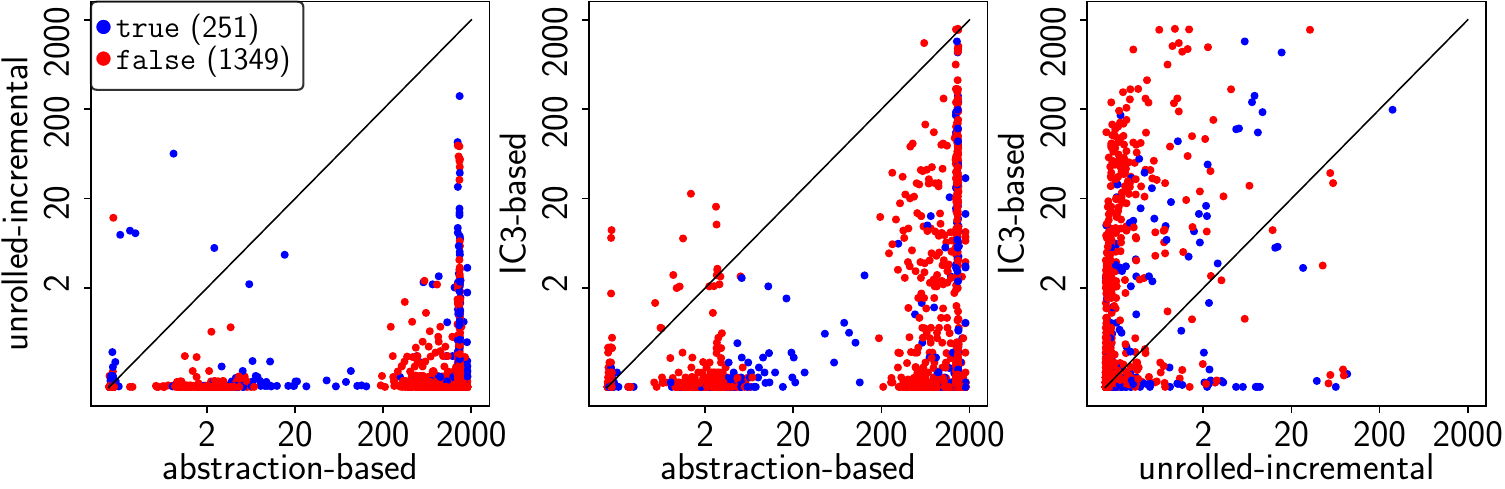}
    \caption{The plots of experiments with TDLTL formulae of size 5.}
  \end{subfigure}
  \begin{subfigure}[!ht]{\textwidth}
    \centering
    \includegraphics[scale=0.55]{plots/abstraction_ic3_unrolled_all_v2.pdf}
    \caption{The plots of experiments with TDLTL formulae of size 10.}
  \end{subfigure}
  \captionsetup{format=hang,width=\linewidth}
    \caption{Comparison of abstraction-based, unrolled-incremental and IC3-based algorithms. 
    }
    \label{BMC_plot1}
\end{figure}
}

The plots in \autoref{BMC_plot1} also indicate that the unrolled-incremental algorithm is more efficient than the IC3-based technique. To shed light on this finding, we then compare the performance of IC3-based technique with SMT-based incremental (unrolled and initialised) algorithms. As depicted in \autoref{BMC_plot2}, the proposed algorithms outperform the procedure that employs IC3. We recall that, in incremental algorithms, the bound of the counterexample is not increased by one. Hence, they are indeed more effective to find a long counterexample. Furthermore, in each iteration, the search of a counterexample is implemented with a fixed loopback bound. Such a bound is related to the current value of the transient $l$.

\ifthenelse{\boolean{papertwocol}}
{
\begin{figure*}[!ht]
  \centering
  \begin{subfigure}[!ht]{\textwidth}
    \centering
    \includegraphics[scale=0.45]{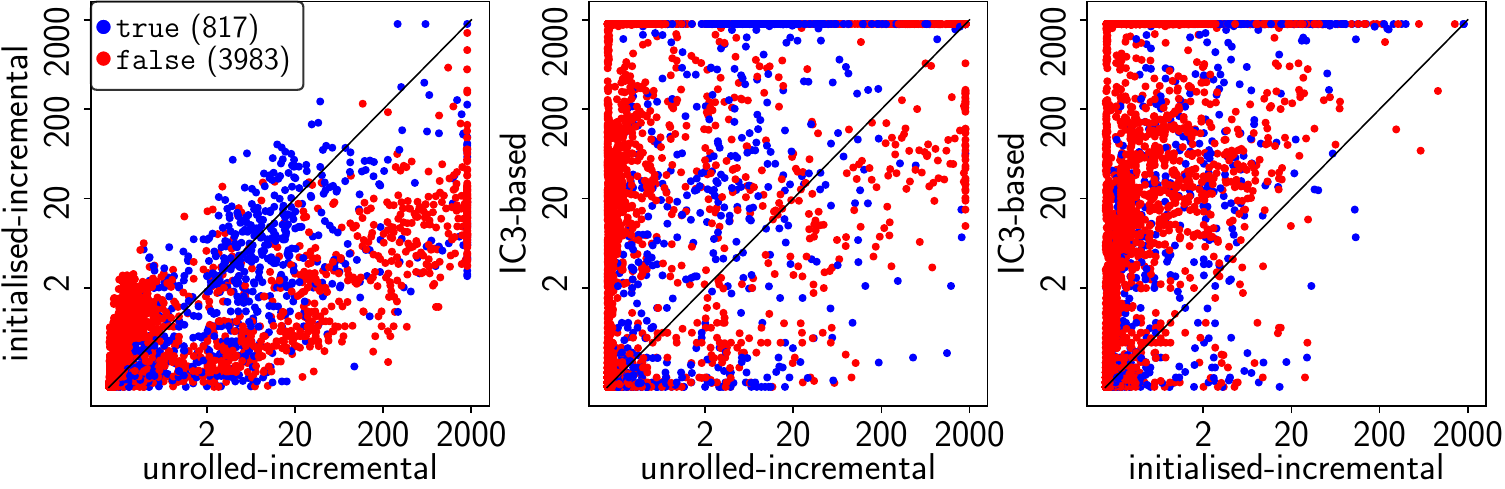}
  \end{subfigure}
  \captionsetup{format=hang,width=0.95\linewidth}
   \caption{Comparison of incremental algorithms and IC3-based techniques.}
    \label{BMC_plot2}
\end{figure*}
\begin{figure*}[!ht]
  \centering
    \includegraphics[scale=0.45]{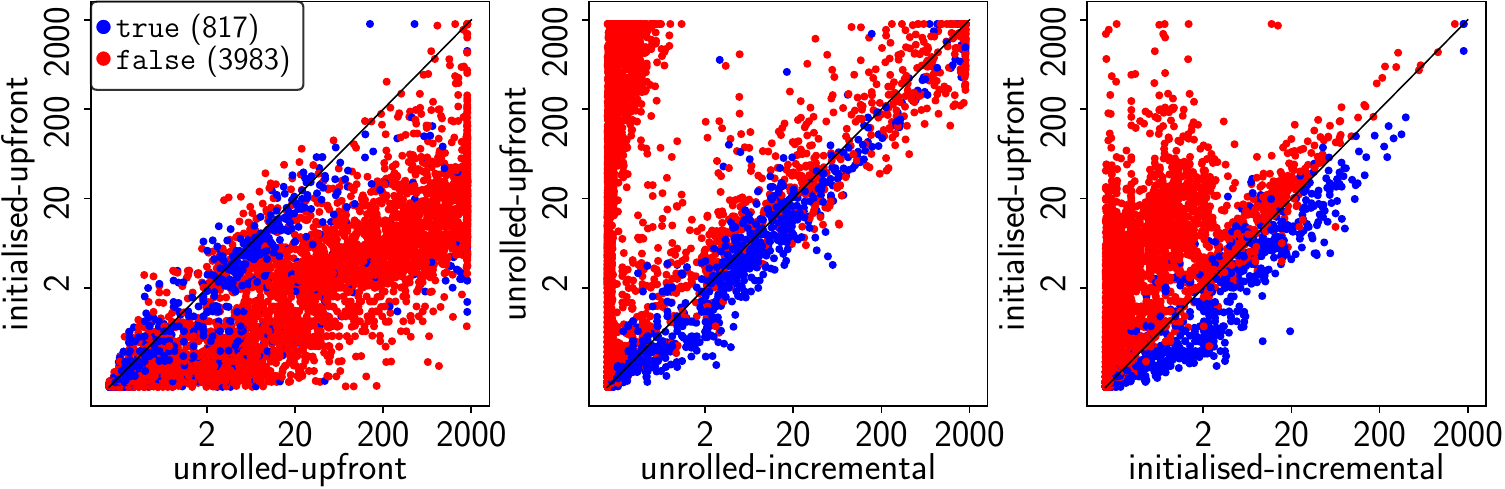}
  \captionsetup{format=hang,width=\linewidth}
   \caption{Comparison of incremental and upfront algorithms.}
    \label{BMC_plot3}
\end{figure*}
}
{
\begin{figure}[!ht]
  \centering
  \begin{subfigure}[!ht]{\textwidth}
    \centering
    \includegraphics[scale=0.55]{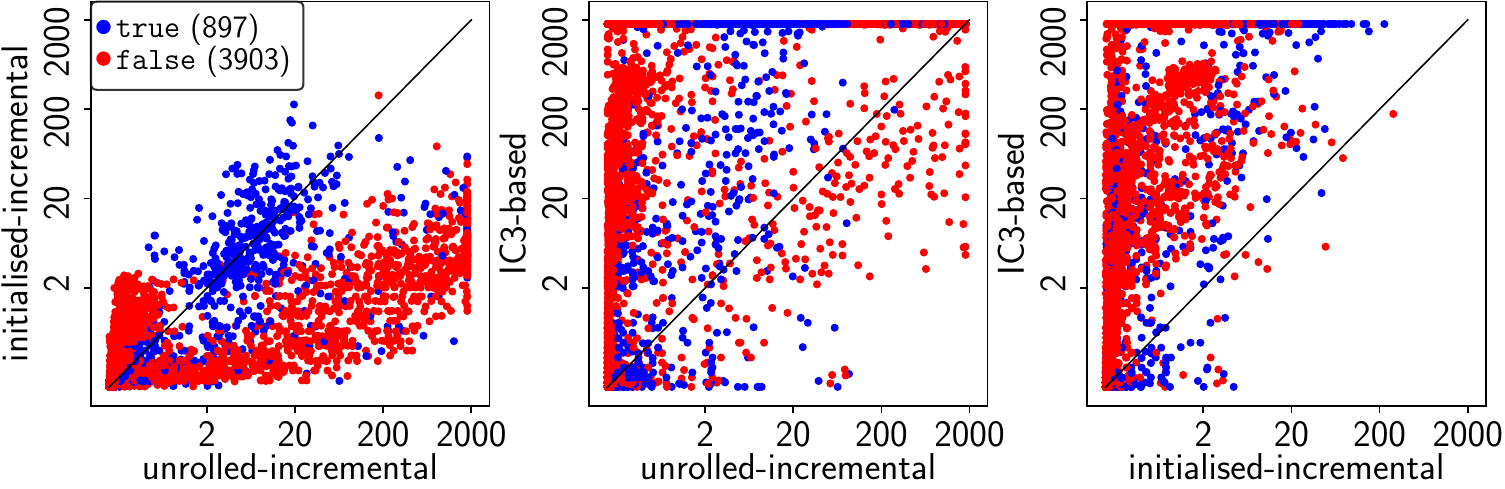}
    \caption{The plots of experiments with TDLTL formulae of size 5.}
  \end{subfigure}
  \begin{subfigure}[!ht]{\textwidth}
    \centering
    \includegraphics[scale=0.55]{plots/ic3_incremental_all_v2.pdf}
    \caption{The plots of experiments with TDLTL formulae of size 10.}
  \end{subfigure}
  \captionsetup{format=hang,width=0.95\linewidth}
   \caption{Comparison of incremental algorithms and IC3-based techniques.}
    \label{BMC_plot2}
\end{figure}
}

Based on our experiments, the dimension of the matrix heavily affects the running time of IC3-based procedures: the number of experiments that yield $\mathtt{timeout}$ increases as the dimension grows. On the other hand, the performance of incremental algorithms is affected by the upper bound of the completeness threshold given by \autoref{CT}, in particular for experiments where the corresponding specification is valid. Between the unrolled-incremental and initialised-incremental algorithms, it seems that the latter one is the winner. We recall that in \autoref{BMC}(b), all TD propositions in \eqref{witness_initial} are initial ones. Therefore, there is only one set of variables that appear in \eqref{witness_initial}. In comparison, there are $k+1$ sets of variables in \eqref{witness}, which correspond to the states of the MPL system \eqref{mpl} from bound 0 until $k$.

We then compare the performance between incremental and upfront algorithms, as shown in \autoref{BMC_plot3}. As expected, upfront procedures are faster than incremental ones when the specification is valid. On the other hand, incremental algorithms are much more efficient when the specification is invalid. This is due to the fact that the counterexample may be found at a smaller bound than the completeness threshold given in \autoref{CT}. As in incremental approaches, the procedure which uses one set of variables (initialised-upfront) is faster than the other one that employs multiple sets of variables (unrolled-upfront).

\ifthenelse{\boolean{papertwocol}}
{
}
{
\begin{figure}[!ht]
  \centering
    \includegraphics[scale=0.55]{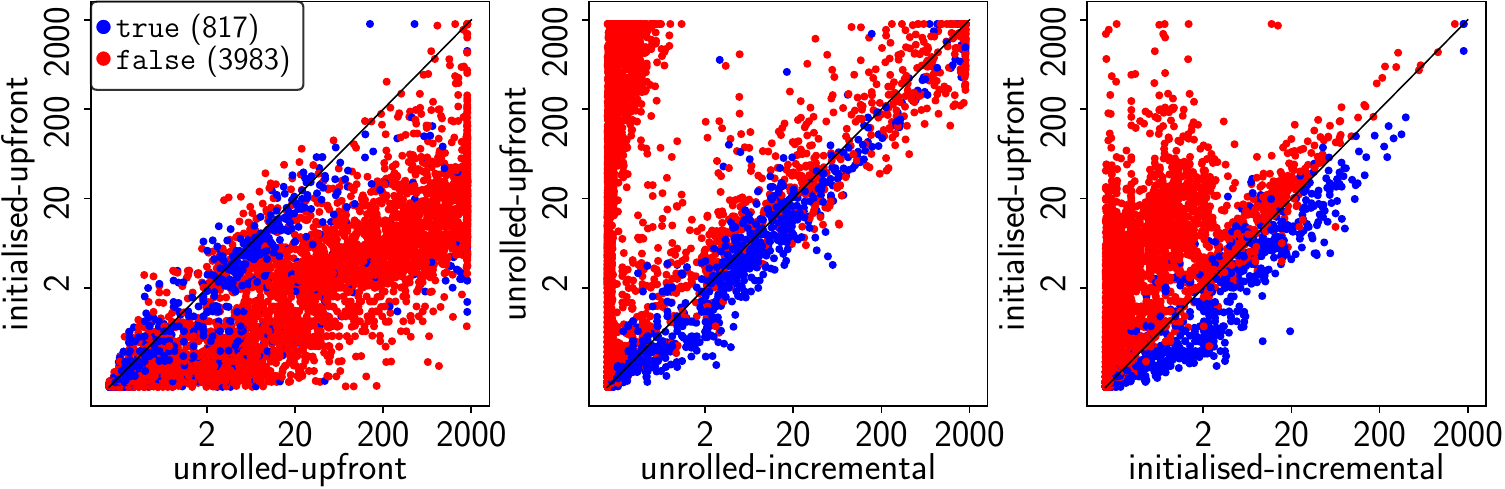}
  \captionsetup{format=hang,width=\linewidth}
   \caption{Comparison of incremental and upfront algorithms.}
    \label{BMC_plot3}
\end{figure}
}

\section{Conclusions and Future Work}
\label{sec-conclusions}


In this paper, we have addressed the problem of automatically analyzing Max-Plus Linear (MPL) systems. 
We have defined the TDLTL logic as a suitable formalism for the specification of temporal properties, and introduced two approaches for cyclicity and transient analysis and for TDTL model checking.
The first approach is based on the direct encoding of MPL into symbolic infinite-state transition systems, and results in algorithms based on general-purpose symbolic model checking.
The second approach is based on the efficient use of Satisfiability Modulo Theories (SMT), and results in optimized algorithms that take into account the specific features of MPL systems. 



As future work, we plan to investigate the case of models with reducible matrices and unbounded transients. 
Furthermore, we will generalize the approach for parametric MPL, where the matrices may contain symbolic expressions, by considering the use of parametric model checking techniques~\cite{parametersynthesis}.

\bibliographystyle{IEEEtran}
\bibliography{references}

\newpage
\appendix
\section{Case Study: The London Underground Network}
\label{sec-case-study}

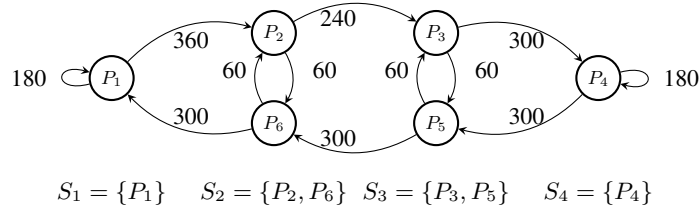
\begin{figure}[!ht]
\centering
\begin{tikzpicture}[node distance=0.75cm and 0.75cm]\footnotesize
\tikzstyle{place}=[circle,thick,draw=black,minimum size=2mm,scale=0.8]
\node[place] (1) {$P_1$};
\node[place] (2) [right =of 1,xshift=1cm,yshift=0.75cm] {$ P_2$};
\node[place] (2a) [right =of 1,xshift=1cm,yshift=-0.75cm] {$ P_6$};
\node[place] (3) [right =of 2,xshift=1cm] {$ P_3$};
\node[place] (3a) [right =of 2a,xshift=1cm] {$P_5$};
\node[place] (4) [right =of 3,xshift=1cm,yshift=-0.75cm] {$ P_{4}$};

\draw[->,>=stealth] (1)[bend left]  to (2);
\draw[->,>=stealth] (2)[bend left]  to (3);
\draw[->,>=stealth] (3)[bend left]   to (4);

\draw[->,>=stealth] (2a)  [bend left] to (1);
\draw[->,>=stealth] (3a)  [bend left] to (2a);
\draw[->,>=stealth] (4)  [bend left] to (3a);
\draw[->,>=stealth] (2a)  [bend left] to (2);
\draw[->,>=stealth] (2)  [bend left] to (2a);
\draw[->,>=stealth] (3a)  [bend left] to (3);
\draw[->,>=stealth] (3)  [bend left] to (3a);
\draw[->,>=stealth] (4)  [loop right] to (4);
\draw[->,>=stealth] (1)  [loop left] to (1);
\node[left=of 1,xshift=0.3cm] {\footnotesize 180};
\node[left=of 2a,xshift=0.3cm,yshift=0.1cm] {\footnotesize 300};
\node[left=of 2,xshift=0.3cm,yshift=-0.1cm] {\footnotesize 360};
\node[left=of 3,xshift=0.1cm,yshift=0.2cm] {\footnotesize 240};
\node[left=of 3a,xshift=0.1cm,yshift=-0.2cm] {\footnotesize 300};
\node[right=of 3,xshift=-0.2cm,yshift=-0.1cm] {\footnotesize 300};
\node[right=of 3a,xshift=-0.2cm,yshift=0.1cm] {\footnotesize 300};
\node[right=of 4,xshift=-0.3cm] {\footnotesize 180};
\node[left=of 2a,xshift=0.8cm,yshift=0.7cm] {\footnotesize 60};
\node[left=of 2a,xshift=2.cm,yshift=0.7cm] {\footnotesize 60};
\node[left=of 3a,xshift=0.8cm,yshift=0.7cm] {\footnotesize 60};
\node[left=of 3a,xshift=2cm,yshift=0.7cm] {\footnotesize 60};
\node[below=of 1,yshift=-0.2cm] {$S_1=\{P_1\}$};
\node[below=of 2,xshift=0cm,yshift=-0.8cm] {$S_2=\{P_2,P_6\}$};
\node[below=of 3,xshift=0cm,yshift=-0.8cm] {$S_3=\{P_3,P_5\}$};
\node[below=of 4,xshift=0cm,yshift=-0.2cm] {$S_4=\{P_4\}$};
\end{tikzpicture}
\caption{A simple network with four stations, six platforms $P_i$, and two routes $P_1\rightarrow P_2\rightarrow P_3\rightarrow P_4$ and $P_4\rightarrow P_5\rightarrow P_6\rightarrow P_1$. Travel/transfer times are in seconds.}
\label{underground}
\end{figure}

We present a max-plus model of eleven lines of the London Underground network. 
Each line serves multiple routes connecting several ``terminus stations'' where a route starts or ends. 
Any station may comprise several ``platforms'' where  passengers wait for incoming trains. 
We call two platforms belonging to the same station ``conjoined''. 
As an illustration, \autoref{underground} shows a network with four stations and six platforms. Here, the edge-weight connecting platforms from different stations represents the travel time of a train (seconds), while the one connecting conjoined platforms is the passenger transfer (walking) time. Self-loops at terminus platforms represent delays for the successive departure of a train: for instance, after a train departs from platform $P_1$ at $m$ time units, then the next departure cannot be earlier than $m+180$ time units. 


The dynamics of the underground model in \autoref{underground} are as follows. Suppose $x_i(k)$ and $t_i(k)$ are the time stamps for the $(k+1)$-th event ``platform $P_i$ is expecting an incoming train'' and the $(k+1)$-th event ``a train departs from $P_i$'', respectively. Note that a train departs from a platform after  travelling from preceding platforms. Therefore, $t_i(k)$ is not necessarily equal to $x_i(k)$. Instead, we have $t_i(k)=x_i(k+d_i)$ where $d_i$ is the smallest \textit{distance} from platform $P_i$ to terminus platforms: if platform $P_i$ is the $k$-th stop from a terminus platform then $d_i=k$. For instance, from model in \autoref{underground}, we have $d_1=d_4=0,d_2=d_5=1$, and $d_3=d_6=2$. The time stamp $x_i(k+1)$ is synchronised over maximization of the time stamps of incoming trains (from adjacent stations) and passengers (from conjoined platforms) to $P_i$. For instance, $x_2(k+1)= \max\{x_{1}(k) + 360,x_{6}(k) + 60\}$.

The overall dynamics of the model in \autoref{underground} can be expressed as follows: 
\ifthenelse{\boolean{papertwocol}}{
\begin{equation*}
\small
\!\begin{bmatrix}
x_1(k+1)\\x_2(k+1)\\x_3(k+1)\\x_4(k+1)\\x_5(k+1)\\x_6(k+1)
\end{bmatrix}\!\!=\!\!
    \begin{bmatrix}
180&~\varepsilon &~\varepsilon &~\varepsilon &~\varepsilon &~300\\
360&~\varepsilon &~\varepsilon &~\varepsilon &~\varepsilon &~60\\
\varepsilon&~240 &~\varepsilon &~\varepsilon &~60&~\varepsilon\\
\varepsilon&~\varepsilon &~300 &~180 &~\varepsilon &~\varepsilon\\
\varepsilon&~\varepsilon &~60 &~300&~\varepsilon &~\varepsilon\\
\varepsilon&~60 &~\varepsilon &~\varepsilon &~300 &~\varepsilon
\end{bmatrix}\!\!\otimes\!\!\begin{bmatrix}
x_1(k)\\x_2(k)\\x_3(k)\\x_4(k)\\x_5(k)\\x_6(k)
\end{bmatrix}
\label{mpl_underground}
\end{equation*}
}{
\begin{equation*}
\begin{bmatrix}
x_1(k+1)\\x_2(k+1)\\x_3(k+1)\\x_4(k+1)\\x_5(k+1)\\x_6(k+1)
\end{bmatrix}=
    \begin{bmatrix}
~180&~\varepsilon &~\varepsilon &~\varepsilon &~\varepsilon &~300\\
~360&~\varepsilon &~\varepsilon &~\varepsilon &~\varepsilon &~60\\
~\varepsilon&~240 &~\varepsilon &~\varepsilon &~60&~\varepsilon\\
~\varepsilon&~\varepsilon &~300 &~180 &~\varepsilon &~\varepsilon\\
~\varepsilon&~\varepsilon &~60 &~300&~\varepsilon &~\varepsilon\\
~\varepsilon&~60 &~\varepsilon &~\varepsilon &~300 &~\varepsilon
\end{bmatrix}\otimes\begin{bmatrix}
x_1(k)\\x_2(k)\\x_3(k)\\x_4(k)\\x_5(k)\\x_6(k)
\end{bmatrix},
\label{mpl_underground}
\end{equation*}
}
with $t_i(k)=x_i(k)$ for $i\in\{1,4\}$, $t_i(k)=x_i(k+1)$ for $i\in\{2,5\}$, and $t_i(k)=x_i(k+2)$ for $i\in\{3,6\}$. 
It is straightforward to check that the graph representation of the underground network, as in \autoref{underground}, is strongly connected. Hence, the underlying max-plus algebraic matrix is irreducible. 
Due to the fact that the initial departures start from the termini $P_1$ and $P_4$, it is conventional to set $x_2(0)=x_3(0)=x_5(0)=x_6(0)=\varepsilon$. Suppose $x_1(0)=x_4(0)=0$, then the sequence of vectors $\textbf{x}(0),\textbf{x}(1),\ldots$ is as follows: 
\ifthenelse{\boolean{papertwocol}}{
\begin{equation*}
\small
\!\begin{bmatrix}
0\\\varepsilon\\\varepsilon\\0\\\varepsilon\\\varepsilon
\end{bmatrix}\!\!,\!
\begin{bmatrix}
180\\360\\\varepsilon\\180\\300\\\varepsilon
\end{bmatrix}\!\!,\!
\begin{bmatrix}
360\\540\\600\\360\\480\\600
\end{bmatrix}\!\!,\!
\begin{bmatrix}
900\\720\\780\\900\\660\\780
\end{bmatrix}\!\!,\!
\begin{bmatrix}
1080\\1260\\960\\1080\\1200\\960
\end{bmatrix}\!\!,\!
\begin{bmatrix}
1260\\1440\\1500\\1260\\1380\\1500
\end{bmatrix}\!\!,\!
\begin{bmatrix}
1800\\1620\\1680\\1800\\1560\\1680
\end{bmatrix}\!\!,\!
\begin{bmatrix}
1980\\2160\\1860\\1980\\2100\\1860
\end{bmatrix}\!\
\ldots
\label{mpl_underground_vector}
\end{equation*}
}{
\begin{equation*}
\begin{bmatrix}
0\\\varepsilon\\\varepsilon\\0\\\varepsilon\\\varepsilon
\end{bmatrix},
\begin{bmatrix}
180\\360\\\varepsilon\\180\\300\\\varepsilon
\end{bmatrix},
\begin{bmatrix}
360\\540\\600\\360\\480\\600
\end{bmatrix},
\begin{bmatrix}
900\\720\\780\\900\\660\\780
\end{bmatrix},
\begin{bmatrix}
1080\\1260\\960\\1080\\1200\\960
\end{bmatrix},
\begin{bmatrix}
1260\\1440\\1500\\1260\\1380\\1500
\end{bmatrix},
\begin{bmatrix}
1800\\1620\\1680\\1800\\1560\\1680
\end{bmatrix},
\begin{bmatrix}
1980\\2160\\1860\\1980\\2100\\1860
\end{bmatrix},
\ldots
\label{mpl_underground_vector}
\end{equation*}
}
Notice that the periodic behavior occurs with transient $l=2$ and cyclicity $c=3$. The corresponding vectors $\textbf{t}(0),\textbf{t}(1),\ldots$ are as follows:
\ifthenelse{\boolean{papertwocol}}{
\[
\small
\!\begin{bmatrix}
0\\360\\600\\0\\300\\600
\end{bmatrix}\!\!,\!
\begin{bmatrix}
180\\540\\780\\180\\480\\780
\end{bmatrix}\!\!,\!
\begin{bmatrix}
360\\720\\960\\360\\660\\960
\end{bmatrix}\!\!,\!
\begin{bmatrix}
900\\1260\\1500\\900\\1200\\1500
\end{bmatrix}\!\!,\!
\begin{bmatrix}
1080\\1440\\1680\\1080\\1380\\1680
\end{bmatrix}\!\!,\!
\begin{bmatrix}
1260\\1620\\1860\\1260\\1560\\1860
\end{bmatrix}\!\!,\!
\begin{bmatrix}
1800\\2160\\2400\\1980\\2100\\2400
\end{bmatrix}\!\!,\!
\begin{bmatrix}
1980\\2340\\2580\\1980\\2280\\2580
\end{bmatrix}
\ldots
\]
}{
\[
\begin{bmatrix}
0\\360\\600\\0\\300\\600
\end{bmatrix},
\begin{bmatrix}
180\\540\\780\\180\\480\\780
\end{bmatrix},
\begin{bmatrix}
360\\720\\960\\360\\660\\960
\end{bmatrix},
\begin{bmatrix}
900\\1260\\1500\\900\\1200\\1500
\end{bmatrix},
\begin{bmatrix}
1080\\1440\\1680\\1080\\1380\\1680
\end{bmatrix},
\begin{bmatrix}
1260\\1620\\1860\\1260\\1560\\1860
\end{bmatrix},
\begin{bmatrix}
1800\\2160\\2400\\1980\\2100\\2400
\end{bmatrix},
\begin{bmatrix}
1980\\2340\\2580\\1980\\2280\\2580
\end{bmatrix},
\ldots
 \]
}
The max-plus model for the London Underground is constructed based on the map and on the actual timetables obtained from Transport for London (TfL), with minor alterations.  
\autoref{LU} presents the the network for each line of London Underground w.r.t. number of routes, number of stations (terminus and non-terminus) and the resulting dimension of the MPL system. 
The self-loop delay at terminus platforms is set to 600 seconds, 
while the transfer time between conjoined platforms is fixed at 60 seconds. 
\begin{table}[tb]
    \centering
    \caption{The London Underground lines considered in the case study.} 
    \begin{tabular}{|c|c|c|c|c|}
    \hline
        Underground line &$\#$routes& $\#$stations& dim \\\hline
         Bakerloo &2&$\{2,23\}$&48\\\hline
         Central &10&$\{4,45\}$&94\\\hline
         District &12&$\{5,54\}$&113\\\hline
         Hammersmith $\&$ City &2&$\{2,27\}$&56\\\hline
         Jubilee &2&$\{2,25\}$&52\\\hline
         Metropolitan &10&$\{5,29\}$&63\\\hline
         Northern &12&$\{4,47\}$&98\\\hline
         Piccadilly &4&$\{3,49\}$&101\\\hline
         Victoria &2&$\{2,14\}$&30\\\hline
         Waterloo $\&$ City &2&$\{2,0\}$&2\\\hline
    \end{tabular}
    \label{LU}
\end{table}


For each model in Table \autoref{LU}, suppose $\textbf{P}$ is the set of platforms. For the sake of simplicity, we write $\textbf{P}=\{1,\ldots,\text{dim}\}$ instead. We can partition $\textbf{P}$ into two sets, a set of terminus platforms $\textbf{T}$ and a set of non-terminus platforms $\textbf{N}$. We consider the following specifications:
\begin{itemize}
    \item[$\bullet$] The absolute difference between departures from terminus platforms is never greater than $300$ seconds: $ t_i(k)-t_j(k)\leq 300~\forall k\geq 0~\text{and}~i,j\in \textbf{T}$.
    Due to the fact that all terminus platforms $i\in \textbf{T}$ have \textit{distance} $d_i=0$, this specification can be expressed as the TDLTL formula $\varphi_1=\bigwedge_{i,j\in \textbf{T}}\ltlG (\texttt{x}_i-\texttt{x}_j\leq 300).$
    \item[$\bullet$] There exits $k>0$ such that the $k$-th departure from terminus platforms happen at the same time. This specification can be expressed as \[\varphi_2=\bigwedge_{i,j\in \textbf{T},i<j} \ltlF (\texttt{x}_i-\texttt{x}_j=0).\]
    \item[$\bullet$] The departures from conjoined platforms are always alternating (i.e. there is always a departure at platform $i$ between two-consecutive departures at platform $j$, or vice versa). This means that for each conjoined platforms $i,j$ with $d_i\geq d_j$ we have either
    \[\forall k\geq 0.~ \texttt{x}_j(d_i+k) < \texttt{x}_i(d_i+k)<\texttt{x}_j(d_i+k+1),\]
    or
    \[\forall k\geq 0.~ \texttt{x}_i(d_i+k) < \texttt{x}_j(d_i+k)<\texttt{x}_i(d_i+k+1).\]
    This specification can be expressed as the TDLTL formula \[\varphi_3=\mathop{\bigwedge_{(i,j)~\text{is}}}_{\text{conjoined}}
    \ltlG (p_{ij} \wedge q_{ji}) \vee \ltlG (p_{ji}\wedge q_{ij}),\]
    where $p_{ij}\!=\!\texttt{x}_i^{(d_i)}-\texttt{x}_j^{(d_i)}\!>\!0$ and $q_{ij}\!=\!\texttt{x}_i^{(d_i+1)}-\texttt{x}_j^{(d_i)}\!>\!0$.
\end{itemize}

The results are presented in Table~\ref{LU2}. For each line of the London Underground, the set of initial conditions is expressed as the LRA formula $X=\bigwedge_{i\in \textbf{T}} 0\leq \texttt{x}_i\leq 300$. \autoref{LU2} presents the results of the experiments, obtained by applying the encoding in  \eqref{witness_initial}. For each line, $\varphi_1$ is valid while $\varphi_2$ is invalid. Instead, the specification $\varphi_3$ is satisfied only for the Bakerloo and the Victoria lines. We then verify the satisfaction of $\neg \varphi_2$ and $\neg \varphi_3$. Although $\varphi_2$ is not valid, from the table one can conclude that there exists a sequence of departure vectors that satisfies $\varphi_2$ for all lines.  The same conclusion holds for $\varphi_3$ (note that $\varphi_3$ is not applicable to the Waterloo $\&$ City line, since it does not have any conjoined platform - hence NA).

\begin{table*}[!ht]
    \centering
    \caption{Outcomes of the verification problems. Running times are in seconds.}
    \resizebox{\textwidth}{!}{
    \begin{tabular}{|c|c|c|c|c|c|c|c|c|c|c|c|}
    \hline
        line &\multicolumn{2}{c|}{$\varphi_1$}&\multicolumn{2}{c|}{$\varphi_2$}&\multicolumn{2}{c|}{$\neg \varphi_2$}&\multicolumn{2}{c|}{$\varphi_3$}&\multicolumn{2}{c|}{$\neg \varphi_3$}\\\hline
         Bakerloo &$\texttt{true}$&1.07&$\texttt{false}$&0.52&$\texttt{false}$&0.53&$\texttt{true}$&12.10 &$\texttt{false}$ &5.85\\\hline
         Central &$\texttt{true}$&32.40&$\texttt{false}$&18.15&$\texttt{false}$&18.16&$\texttt{false}$&200.89&$\texttt{false}$&201.45\\\hline
         District&$\texttt{true}$&120.61&$\texttt{false}$&64.82&$\texttt{false}$&64.83&$\texttt{false}$&636.35&$\texttt{false}$&638.75\\\hline
         Hammersmith $\&$ City& $\texttt{true}$&1.70&$\texttt{false}$&0.98&$\texttt{false}$&0.98&$\texttt{false}$&12.93&$\texttt{false}$&13.01\\\hline
         Jubilee &$\texttt{true}$&0.72&$\texttt{false}$&0.73&$\texttt{false}$&0.73&$\texttt{false}$&8.81&$\texttt{false}$&8.89\\\hline
         Metropolitan&$\texttt{true}$&9.70&$\texttt{false}$&5.10&$\texttt{false}$&5.11&$\texttt{false}$&87.69&$\texttt{false}$&40.70\\\hline
         Northern&$\texttt{true}$&34.29&$\texttt{false}$&18.70&$\texttt{false}$&18.71&$\texttt{false}$&253.77&$\texttt{false}$&253.36\\\hline
         Piccadilly&$\texttt{true}$&16.25&$\texttt{false}$&16.84&$\texttt{false}$&16.83&$\texttt{false}$&303.12&$\texttt{false}$&303.58\\\hline
         Victoria&$\texttt{true}$&1.08&$\texttt{false}$&1.00&$\texttt{false}$&1.01&$\texttt{true}$&0.67&$\texttt{false}$&0.68\\\hline
         Waterloo $\&$ City& $\texttt{true}$&0.002&$\texttt{false}$&0.002&$\texttt{false}$&0.002&NA&NA&NA&NA\\\hline
    \end{tabular}
    }   
    \label{LU2}
\end{table*}

\bigskip

\newpage
\section{Additional Experiments for Computing Transient and Cyclicity}
\label{sec-more experiments}
This appendix presents the result of experiments to compute the transient and cyclicity for dimension $n\in \{4,6,8,10,20,30\}$ and several values for $m$. In general, the overall outcome is similar to experiments in \autoref{benchmark}: the SMT-based algorithm outperforms the rest while the STS-based technique is the least efficient. 

\begin{figure}[!ht]
  \begin{subfigure}[t]{0.5\textwidth}
  \centering
  \includegraphics[scale=0.7]{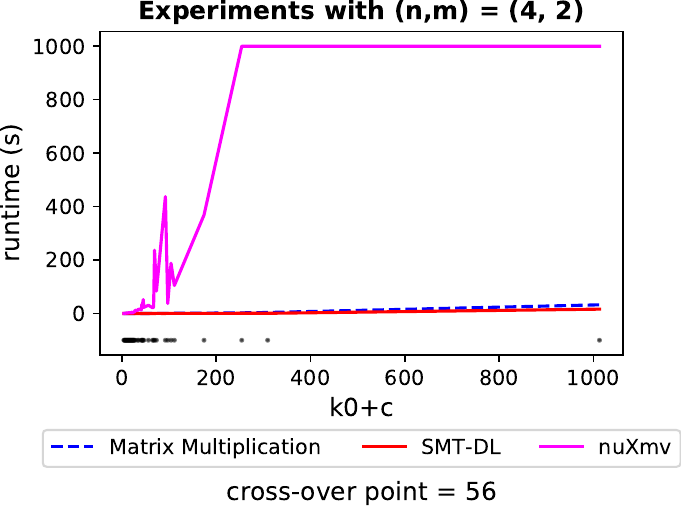}
\end{subfigure}
\begin{subfigure}[t]{0.5\textwidth}
  \centering
  \includegraphics[scale=0.7]{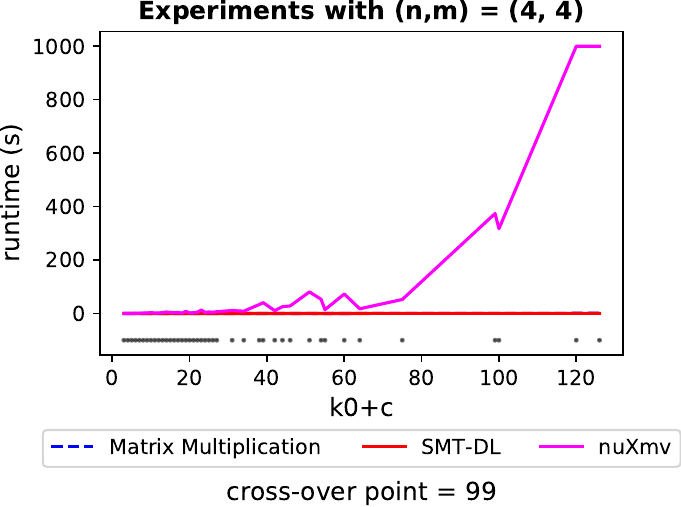}
\end{subfigure}
\vspace*{2ex}\\
  \begin{subfigure}[t]{0.5\textwidth}
  \centering
  \includegraphics[scale=0.7]{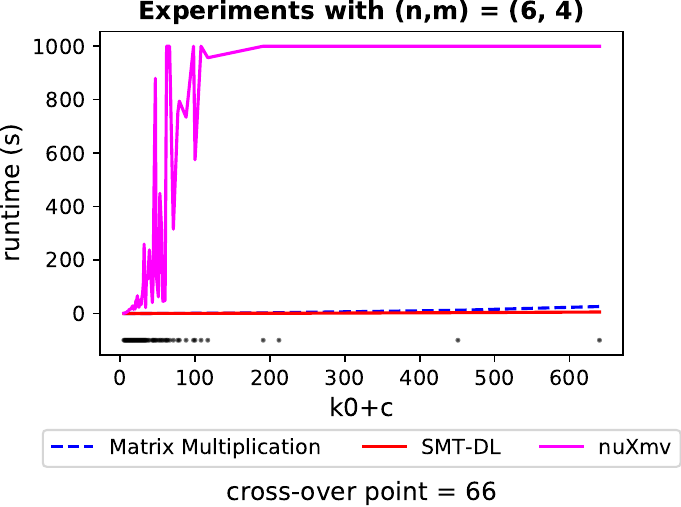}
\end{subfigure}
\begin{subfigure}[t]{0.5\textwidth}
  \centering
  \includegraphics[scale=0.7]{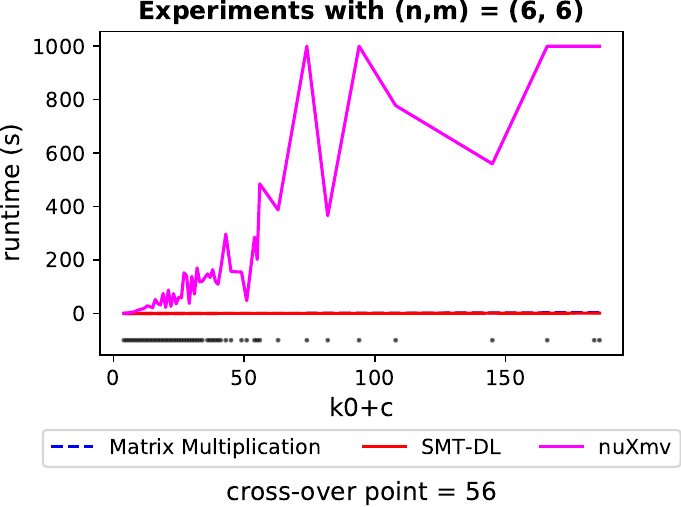}
\end{subfigure}
\end{figure}
\begin{figure}[!ht]
\begin{subfigure}[t]{0.5\textwidth}
\centering
\includegraphics[scale=0.7]{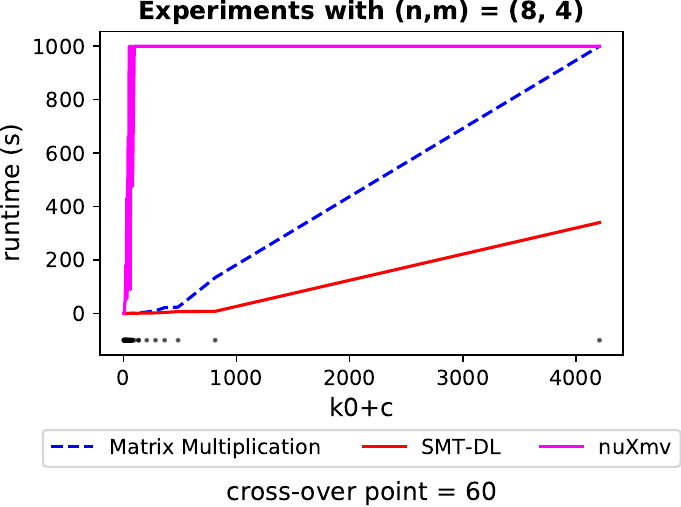}
\end{subfigure}
\begin{subfigure}[t]{0.5\textwidth}
\centering
\includegraphics[scale=0.7]{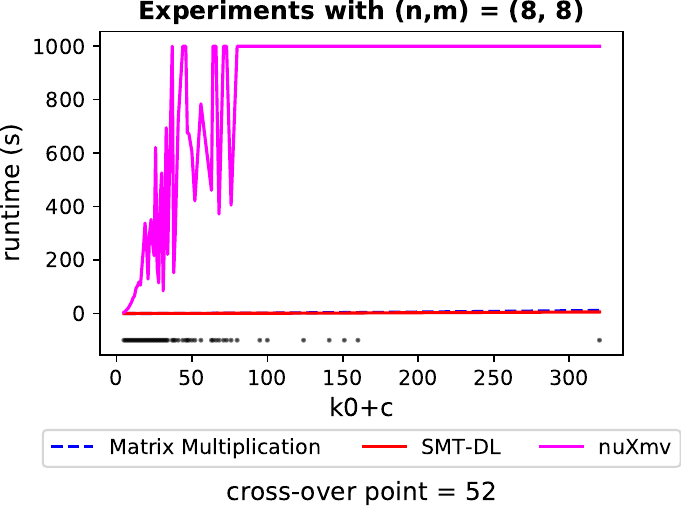}
\end{subfigure}
\vspace*{2ex}\\
\begin{subfigure}[t]{0.5\textwidth}
\centering
\includegraphics[scale=0.7]{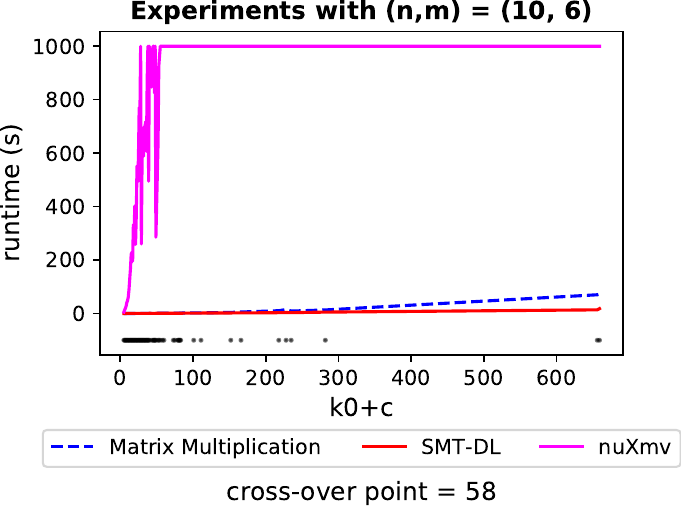}
\end{subfigure}
\begin{subfigure}[t]{0.5\textwidth}
\centering
\includegraphics[scale=0.7]{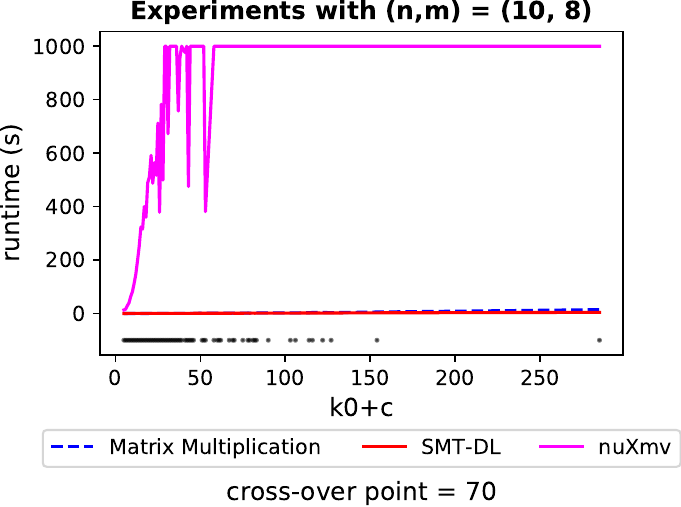}
\end{subfigure}
\vspace*{2ex}\\
\begin{subfigure}[t]{0.5\textwidth}
\centering
\includegraphics[scale=0.7]{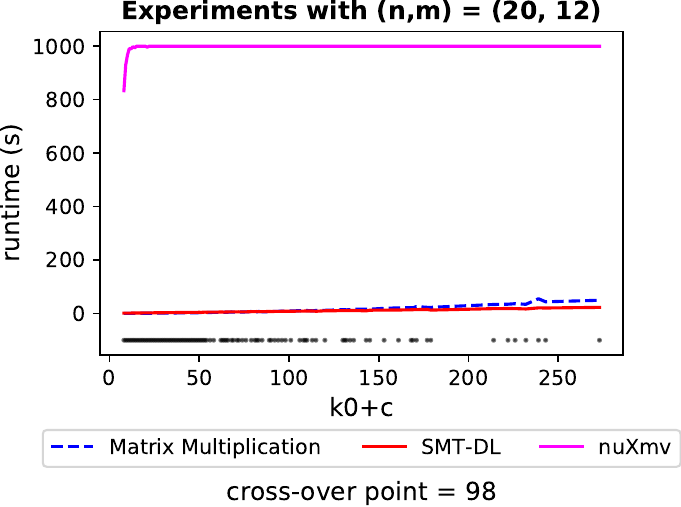}
\end{subfigure}
\begin{subfigure}[t]{0.5\textwidth}
\centering
\includegraphics[scale=0.7]{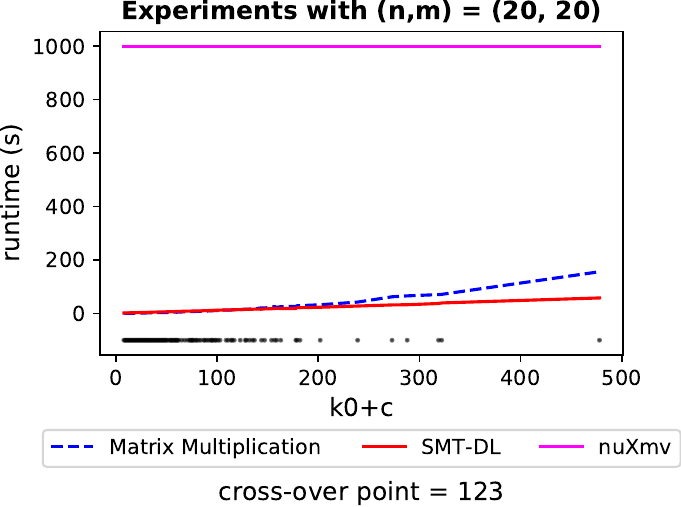}
\end{subfigure}
\end{figure}

\begin{figure}[!ht]
\begin{subfigure}[t]{0.5\textwidth}
\centering
\includegraphics[scale=0.7]{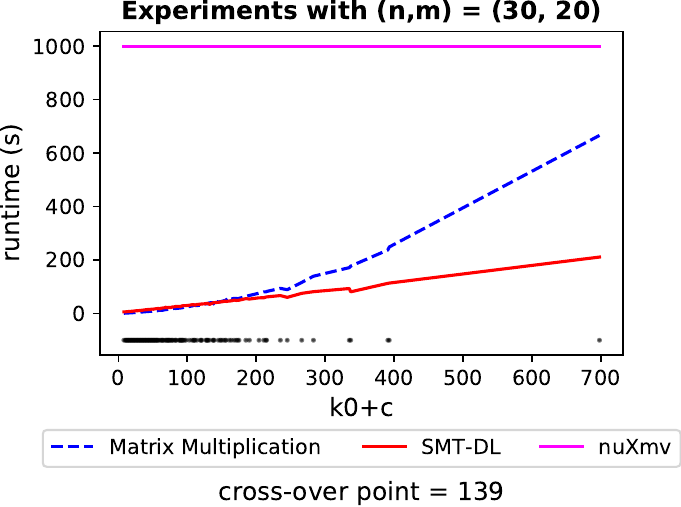}
\end{subfigure}
\begin{subfigure}[t]{0.5\textwidth}
\centering
\includegraphics[scale=0.7]{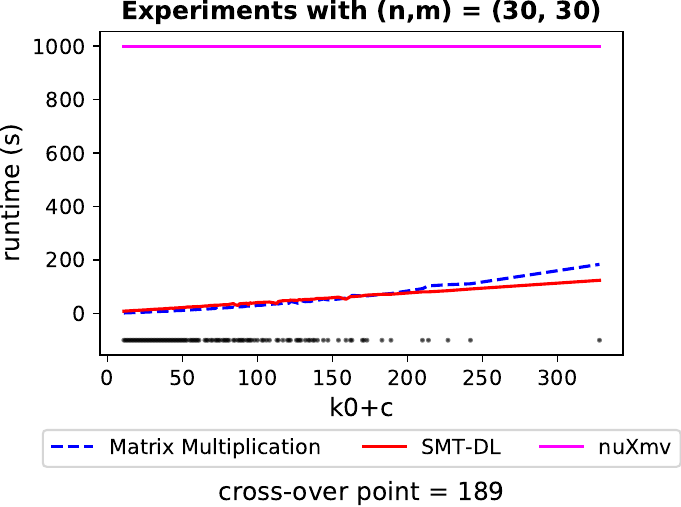}
\end{subfigure}
\end{figure}

\end{document}